\documentstyle[11pt,aaspp4]{article}
\begin{document}

\title{DIRECT Distances to Nearby Galaxies Using Detached Eclipsing
Binaries and Cepheids.  VII. Additional Variables in the Field M33A
Discovered with Image Subtraction\footnote{Based on observations
obtained with the 2.1m telescope at the Kitt Peak National
Observatory.}}

\author{B. J. Mochejska, J. Kaluzny}
\affil{Copernicus Astronomical Center, Bartycka 18, 00-716 Warszawa}
\affil{\tt e-mail: mochejsk@camk.edu.pl, jka@camk.edu.pl}
\author{K. Z. Stanek\altaffilmark{2}, D. D. Sasselov\altaffilmark{3} 
\& A. H. Szentgyorgyi}
\affil{Harvard-Smithsonian Center for Astrophysics, 60 Garden St.,
Cambridge, MA~02138}
\affil{\tt e-mail: kstanek@cfa.harvard.edu, sasselov@cfa.harvard.edu
aszentgyorgyi@cfa.harvard.edu}
\altaffiltext{2}{Hubble Fellow.}
\altaffiltext{3}{Alfred P. Sloan Research Fellow.}

\begin{abstract}

DIRECT is a project to obtain directly the distances to two Local
Group galaxies, M31 and M33, which occupy a crucial position near the
bottom of the cosmological distance ladder.

As the first step of the DIRECT project we have searched for detached
eclipsing binaries (DEBs) and new Cepheids in the M31 and M33 galaxies
with 1m-class telescopes. In this seventh paper we present a catalog
of variable stars discovered in the data from the followup
observations of DEB system D33J013346.2+304439.9 in field M33A
$[(\alpha,\delta)= (23.\!\!\arcdeg55, 30.\!\!\arcdeg72), {\rm
J2000.0}]$, collected with the Kitt Peak National Observatory 2.1m
telescope. In our search covering an area of $108\arcmin^2$ we have
found 434 variable stars: 63 eclipsing binaries, 305 Cepheids, and 66
other periodic, possible long period or non-periodic variables. Of
these variables 280 are newly discovered, mainly short-period and/or
faint Cepheids. Their light curves were extracted using the ISIS image
subtraction package. For 85\% of the variables we present light curves
in standard $V$ and $B$ magnitudes, with the remaining 15\% expressed
in units of differential flux.

We have discovered a population of first overtone Cepheid candidates
and for eight of them we present strong arguments in favor of this
interpretation. 

We also report on the detection of a non-linearity in the KPNO T2KA
and T1KA cameras.

The catalog of variables, as well as their photometry (about
$7.8\times 10^4$ BV measurements) and finding charts, is available
electronically via {\tt anonymous ftp} and the {\tt World Wide Web}.
The complete set of the CCD frames is available upon request.
\end{abstract}

\section{Introduction}

Starting in 1996 we undertook a long term project, DIRECT
(i.e. ``direct distances''), to obtain the distances to two important
galaxies in the cosmological distance ladder, M31 and M33. These
``direct'' distances will be obtained by determining the distance of
Cepheids using the Baade-Wesselink method and by measuring the
absolute distance to detached eclipsing binaries (DEBs).  While the
cosmological distance scale has been the subject of numerous recent
observation campaigns, especially those enabled by the Hubble Space
Telescope (HST) and massive variability studies of the Magellanic
clouds, M33 has not been re-surveyed since the photographic
survey of Kinman, Mould \& Wood (1987).

M31 and M33 are the stepping stones to most of our current effort to
understand the evolving universe at large scales.  First, they are
essential to the calibration of the extragalactic distance
scale. Second, they constrain population synthesis models for early
galaxy formation and evolution and provide the stellar luminosity
calibration. There is one simple requirement for all this---accurate
distances. These distances are now known to no better than 10-15\%, as
there are discrepancies of $0.2-0.3\;{\rm mag}$ between various
distance indicators (e.g.~Huterer, Sasselov \& Schechter 1995; Holland
1998; Stanek \& Garnavich 1998).

DEBs have the potential to establish distances to M31 and M33 with an
unprecedented accuracy of better than 5\% and possibly to better than
1\%. Detached eclipsing binaries (for reviews see Andersen 1991;
Paczy\'nski 1997) offer a single step distance determination to nearby
galaxies and may therefore provide an accurate zero point calibration
of various distance indicators -- a major step towards very accurate
determination of the Hubble constant, presently an important but
daunting problem for astrophysicists. DEBs have been recently used to
obtain accurate distance estimate to the Large Magellanic Cloud
(e.g. Guinan et al.~1998; Udalski et al.~1998).

The detached eclipsing binaries have yet to be used as distance
indicators to M31 and M33. According to Hilditch (1996) there was only
{\em one} eclipsing binary of any kind known in M33 (Hubble 1926). The
recent availability of large-format CCD detectors and inexpensive CPUs
has made it possible to organize a massive search for periodic
variables, which will produce a handful of good DEB candidates. These
can then be spectroscopically followed-up with the powerful new 6.5-10
meter telescopes.

The study of Cepheids in M33 has a venerable history (Hubble 1926).
Freedman, Wilson \& Madore (1991) obtained multi-band CCD photometry
of some of the Cepheids discovered in photographic surveys, to build a
period-luminosity relations in M33. However, the sparse photometry and
the small sample (11 Cepheids) do not provide a good basis for
obtaining direct Baade-Wesselink distances (see, e.g., Krockenberger,
Sasselov \& Noyes 1997) to Cepheids---the need for new digital
photometry has been long overdue.

As the first step of the DIRECT project we have searched for DEBs and
new Cepheids in the M31 and M33 galaxies. We have analyzed five
$11\arcmin\times11\arcmin$ fields in M31, A-D and F (Kaluzny et
al. 1998, 1999; Mochejska et al. 1999; Stanek et al. 1998, 1999;
hereafter Papers I, IV, V, II, III). A total of 410 variables, mostly
new, were found: 48 eclipsing binaries, 206 Cepheids and 156 other
periodic, possible long-period or non-periodic variables. We have also
analyzed two fields in M33, A and B (Macri et al. 2001a; hereafter
Paper VI) and found 544 variables: 47 eclipsing binaries, 251 Cepheids
and 246 other variables.

As a second step, we started followup observations of selected DEBs
in both the M31 and M33 galaxies with bigger telescopes in order to 
construct more precise and well sampled light curves for them.

In this paper, seventh in the series, we present a catalog of variable
stars found in the same field as the detached eclipsing binary
D33J013346.2+304439.9 using followup observations collected at the
Kitt Peak National Observatory 2.1m telescope. The paper is organized
as follows: Section 2 provides a description of the observations. The
data reduction procedure is outlined in Section 3. The catalog of
variable stars is presented in Section 4, followed by its brief
discussion in Section 5. Section 6 deals with the first overtone
Cepheid candidates. The concluding remarks are stated in Section 7.

\section{Observations}

\begin{figure}[t]
\plotfiddle{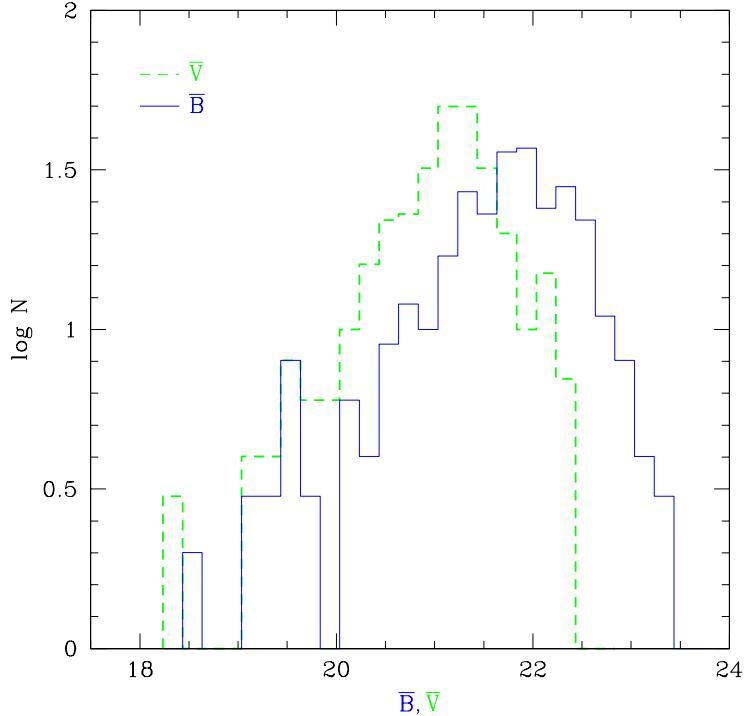}{8.8cm}{0}{50}{50}{-150}{-85}
\caption{Distributions in B (continuous line) and V (dashed line)
of variable stars in the field M33A.}
\label{fig:dist}
\end{figure}

The data discussed in this paper was obtained at the Kitt Peak
National Observatory\footnote{Kitt Peak National Observatory is a
division of NOAO, which are operated by the Association of
Universities for Research in Astronomy, Inc. under cooperative
agreement with the National Science Foundation.} 2.1m telescope
equipped with a Tektronix $2048\times2048$ CCD (T2KA camera) having a
pixel scale $0.305\arcsec/pixel$ during two separate runs, from
September 29th to October 5th, 1999 and from November 1st to 7th,
1999.  The primary observing targets were three detached eclipsing
binaries, one in each of the fields M33A, M33B and M31A, discovered
previously as part of the DIRECT project (Papers VI and II). For field
M33A we collected $158\times600s$ exposures in the $V$ filter and
$62\times600s$ in the $B$ filter.\footnote{The complete list of
exposures for this field and related data files are available through
{\tt anonymous ftp} on {\tt cfa-ftp.harvard.edu}, in {\tt
pub/kstanek/DIRECT} directory. Please retrieve the {\tt README} file
for instructions.  Additional information on the DIRECT project is
available through the {\tt WWW} at {\tt
http://cfa-www.harvard.edu/\~\/kstanek/DIRECT/}.} The exposure times
varied slightly to compensate for the changes of seeing
conditions. The typical seeing was $1.\!\!\arcsec4$. The field was
observed through airmass ranging from 1 to 1.9, with the average at
1.2. The completeness of our data starts to drop rapidly at about 21.2
mag in $V$ and 21.8 mag in $B$, judging from the magnitude
distributions of the variable stars (Fig. \ref{fig:dist}).

\section{Data reduction}
\subsection{Correction for the non-linearity of the T2KA camera}

During the second night of the first observing run we compared the
photometry from two images of NGC 7789 with differing exposure times
and found that the camera had a non-linear response. To lessen the
impact of non-linearity we reduced the gain setting from 5.1 to 3.6
$e^-$/ADU.

Several sets of calibration frames were collected at both gain
settings for the purpose of correcting the data for the
non-linearity. Each set consisted of a series of dome flats with
intensities from $\sim500$ ADU to $\sim60000$ ADU. After subtracting
the readout bias under IRAF\footnote{IRAF is distributed by the
National Optical Astronomy Observatories, which are operated by the
Association of Universities for Research in Astronomy, Inc., under
cooperative agreement with the NSF.} the mode intensity $I_i$ (in ADU)
was computed for each calibration frame.

We adopted the correction function for the non-linearity in the same
form as in the IRAF task IRLINCOR:
\begin{equation}
I_e = I_i \cdot (c_1 + c_2 \cdot\frac{I_i}{32767} 
+ c_3 \cdot (\frac{I_i}{32767})^2)
\label{eq:lincor} 
\end{equation}
where $I_i$ is the mode observed intensity on the calibration frame,
$I_e$ the intensity corrected to the expected level and the
coefficients $c_1, c_2, c_3$ are to be determined. The exposure time,
$t$ corrected for the shutter speed (0.1 s), was used to calculate
the expected intensity $I_e$ from the measured intensity level $I_i$:
\begin{eqnarray*}
I_e = I_i \cdot \frac{t}{t_0}
\end{eqnarray*}
By definition the detector response was assumed to be linear at the
intensity level of the image with the shortest exposure time $t_0$.
The coefficients $c_1$, $c_2$ and $c_3$ were determined from a
least-squares fit of Eq. \ref{eq:lincor} to the measured and expected
intensity levels $I_i$ and $I_e$. We list them in Tab. \ref{tab:coeff}
for the two gain settings used.

\begin{figure}[t]
\plotfiddle{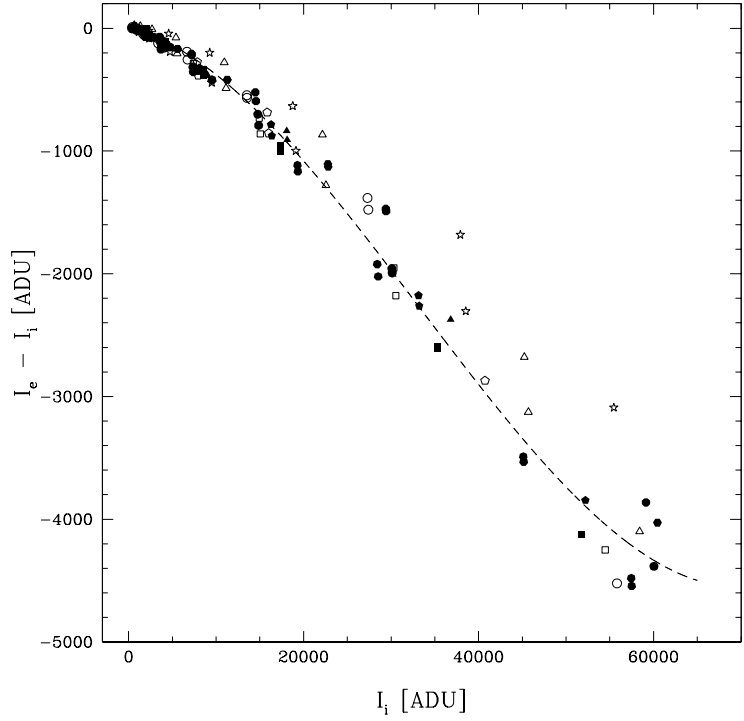}{8.8cm}{0}{50}{50}{-150}{-85}
\caption{The difference between the expected intensity $I_e$ and the
measured intensity $I_i$ as a function of the measured intensity. Each
point corresponds to a calibration exposure. Exposures from the same
set are marked with like symbols. Only the points indicated by filled
symbols were used to fit the correction function (dashed line).}
\label{fig:lincor}
\end{figure} 

\begin{small}
\tablenum{1}
\begin{planotable}{lrrr}
\tablewidth{25pc}
\tablecaption{\sc Coefficients for the Non-linearity Correction Function}
\tablehead{ \colhead{gain ($e^-$/ADU)} & \colhead{$c_1$} & \colhead{$c_2$}
& \colhead{$c_3$}}
\startdata
  3.6 & 0.983282 & $-$0.0765595 &    0.0252555 \nl 
  5.1 & 0.967008 & $-$0.0354981 & $-$0.0314427 \nl
\enddata
\label{tab:coeff}
\end{planotable}
\end{small}

Figure \ref{fig:lincor} shows the difference between the expected
intensity $I_e$ and the measured intensity $I_i$ as a function of the
measured intensity for gain 3.6$e^-$/ADU. Each point corresponds to a
calibration exposure. Exposures from the same set are marked with like
symbols. The points exhibit substantial scatter between the sets and
in some cases within the set (ie. open stars and open
triangles). Several fits were made to different combinations of
sets. The derived coefficients were then tested on pairs of frames of
the same object with exposure times differing by a factor of at least
10. The fit corresponding to the best coefficients is shown with a
dotted line on Fig. \ref{fig:lincor}. The calibration frames used in
the fit are indicated by filled symbols.

\begin{figure}[htp]
\plotfiddle{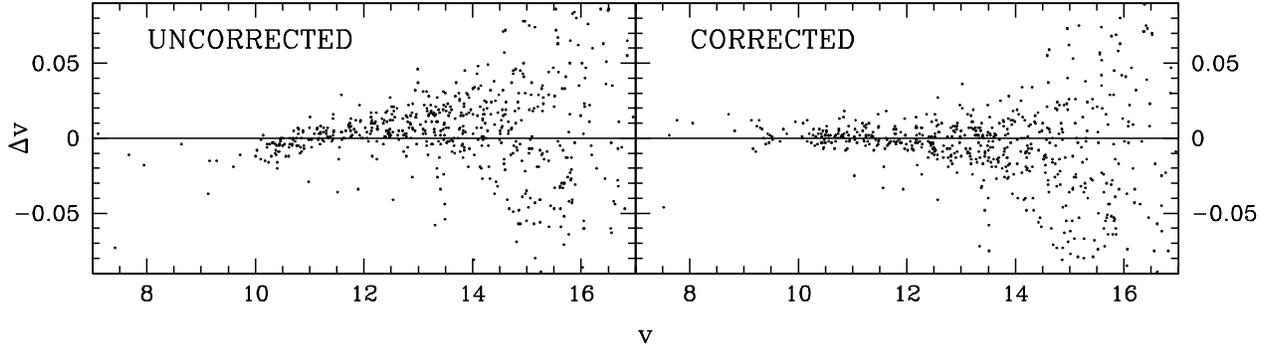}{4.0cm}{0}{85}{85}{-265}{-460}
\caption{A comparison between the T2KA photometry obtained
from a 100s and a 10s exposure of the open cluster NGC 7789. Left
panel: No correction was applied. Right panel: The exposures were 
corrected with a third order polynomial.}
\label{fig:lincmp}
\end{figure}

A comparison between the photometry derived from a 100s and a 10s
exposure of the NGC 7789 open cluster is presented in
Fig.\ref{fig:lincmp}. The left panel shows a comparison for the
uncorrected exposures and the right panel for the corrected ones. A
significant improvement is seen between the uncorrected and corrected
frames. A close inspection of the right panel of the figure will
reveal that the non-linearity has not been removed completely.

The standard preliminary processing of the science data was performed
with the routines in the IRAF CCDPROC package. The non-linearity was
corrected with the IRLINCOR task using the derived coefficients.

\subsection{Nonlinearity of the T1KA CCD}

\begin{figure}[htp]
\plotfiddle{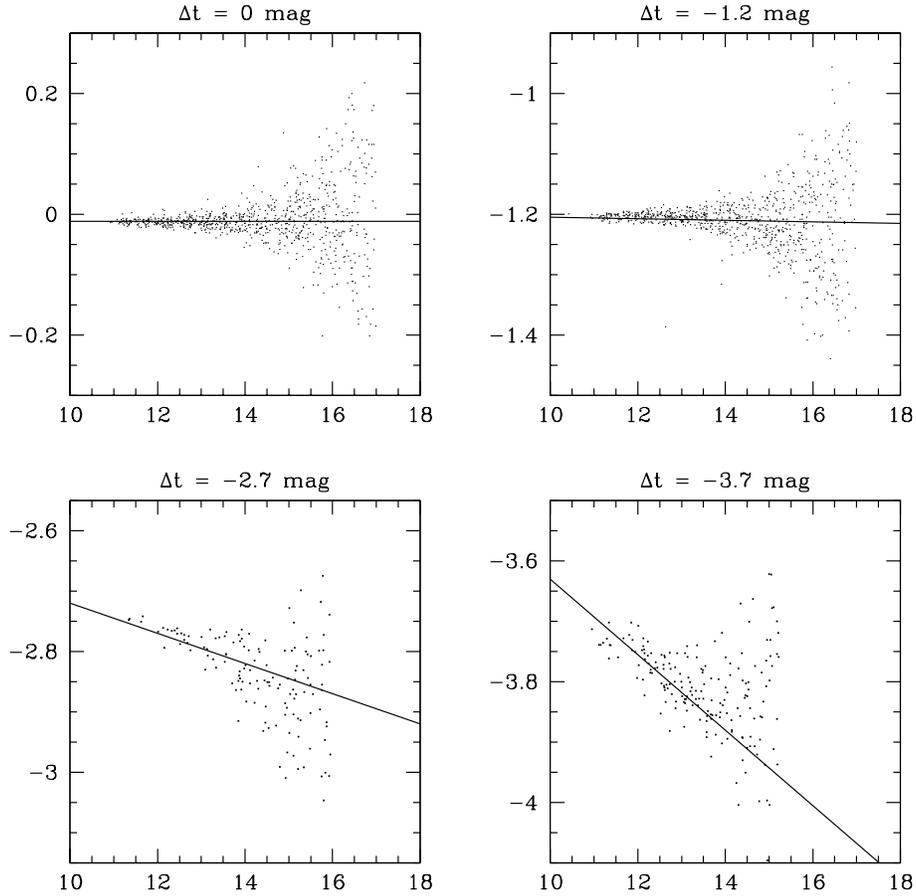}{11.0cm}{0}{65}{65}{-200}{-120}
\caption{A comparison between the T1KA photometry of the open cluster Be 32
obtained from exposures of varying lengths. Going from the upper left
to the lower right panel the photometry from a 180 s exposure is compared
with that of: 180 s, 60 s, 15 s and 6 s.}
\label{fig:1k}
\end{figure}

During an observing run in October 1998 at the KPNO 2.1 meter
telescope we have collected data using the T1KA $1024\times1024$
CCD. We have compared a series of images of an open cluster, Be 32,
which were obtained one after another in photometric conditions
(Fig. \ref{fig:1k}). The images vary only in exposure times (starting
at 6 seconds and going up to 180 seconds). We have found that the
magnitudes (obtained via point spread function fitting + aperture
correction) do not differ by the expected factor of 2.5 log (time),
but rather show increasing deviations from the expected scaling as the
difference in exposure time gets larger. Furthermore, the relations
show strong non-linearity, which increases with the difference in
exposure time.  This fact was not known to us during the time of
observations and no calibration frames were collected to correct the
data. We provide this information to warn the astronomical community
of the problem we have encountered with this camera.

\subsection{Photometry and Selection of Variables}
The photometry for the variable stars was extracted using the ISIS
image subtraction package (Alard \& Lupton 1998, Alard 2000a) from the
$V$ and $B$-band data. 

The ISIS reduction procedure consists of several steps. Initially all
of the frames are transformed to a common coordinate grid. Next a
reference image is created by stacking several frames with the best
seeing. For each frame the reference image is convoluted with a kernel
to match its point spread function (PSF) and then subtracted. On the
subtracted images the constant stars will cancel out and only the
signal from variable stars should remain. A median image is
constructed of all the subtracted images and the variable stars are
identified as bright peaks on it. Finally, profile photometry is
extracted from the subtracted images.

All of the computations were performed with the frames internally
subdivided into four sections (the ISIS parameters sub\_x and sub\_y
set to 2). The differential brightness variations of the background
were fit with a second degree polynomial (deg\_bg=2). A convolution
kernel varying quadratically with position was chosen (deg\_spatial=2).

An image of particularly good quality was selected as the template
frame for the stellar positions. The remaining images were re-mapped
to the template frame coordinate system using a third degree
polynomial transform. During this step an initial rejection of cosmic
rays was also performed. A setting of 1.0 for the cosmic ray threshold
(COSMIC\_THRESH) was used.

A reference image was then constructed from 25 images with the best
seeing. The minimum, maximum and average seeing of these 25 images
were $0.95\arcsec$, $1.23\arcsec$ and $1.11\arcsec$ in $V$ and
$0.93\arcsec$, $1.28\arcsec$ and $1.13\arcsec$ in $B$. The constituent
images were transformed to the same seeing and background level as the
best image (the template in our case) and stacked by taking a median
in each pixel to obtain the reference image virtually free of cosmic
rays.

Image subtraction was then applied to all the frames. For each frame
the reference image was convoluted with a kernel to match it as
closely as possible and then the frame was subtracted from it. As the
flux of the non-variable stars on both images should be almost
identical, such objects will disappear from the subtracted image. The
only remaining signal will come from variable stars.

To obtain a better signal-to-noise ratio a median of the subtracted
images was computed for the purpose of detecting variable stars. The
variables were then identified as bright peaks on this image. In
practice the situation is slightly complicated by the presence of
stars saturated on the reference image, as their profiles were not
matched exactly to those on frames with inferior seeing.  An easy way
to distinguish genuine variables from saturated stars is by their
profile on the subtracted image: it should be smooth for the variables
and usually contains a flat top or a dip at the center for the
saturated stars. The variables were identified visually on the median
subtracted image and their positions recorded with the IMEXAMINE task
under IRAF. The final positions were derived with the {\sc daophot}
FIND subroutine (Stetson 1987). Variables were identified in both
bands. The lists were then merged into a master variable list.

The light curves of the variables (in ADU) were extracted from the
subtracted images. Preliminary periods were found with the ISIS {\it
czerny} subroutine, based on the analysis of variance method, as
described by Schwarzenberg-Czerny (1989), and later refined using
the same algorithm. Stars with chaotic and/or noisy light curves were
rejected from the sample upon visual inspection.

In the next step the light curves were converted into magnitudes.
PSF-fitting photometry was extracted with the {\sc daophot/allstar}
package from the template image following the approach described in
Paper I. Some of the variable stars were not detected on
the template, in most cases because of their faintness or
crowding. As light curves of these stars could not be converted
to magnitudes, they were left expressed in counts.

The template instrumental magnitude $m_{tpl}$ of each variable was
converted into counts $c_{tpl}$, using the {\sc Allstar} zeropoint of
25 mag. The light curve was then converted point by point to
magnitudes $m_i$, by computing the total flux $c_i$ for the given
epoch as the sum of the counts on the template $c_{tpl}$ and the
counts on the subtracted template image $\Delta c_{tpl} =
c_{ref}-c_{tpl}$ decreased by the counts corresponding to the
subtracted image $\Delta c_i=c_{ref}-c_i$.
\begin{equation}
c_i = c_{tpl} + \Delta c_{tpl} - \Delta c_i
\end{equation}
The flux was converted to instrumental magnitudes, using the same zero
point as above. To convert the photometric error $\sigma_{c,i}$ from
counts to $\sigma_{m,i}$ magnitudes, we used the following relation:
\begin{equation}
\sigma_{m,i} = -2.5 \log (\frac{c_i}{c_i+\sigma_{c,i}})
\end{equation}

\subsection{Photometric Calibration and Astrometry}
On the nights of October 3/4 and November 3/4 1999 we observed a total
of 75 stars in 13 Landolt (1992) fields. We attempted to derive a
calibration from these exposures, corrected for non-linearity. The o-c
residuals showed a clear dependence on magnitude, especially in the
$V$-band, where they exhibited a very similar slope on both
nights. Several new fits to Eq. 1 with different combinations of data
sets were tried. Those that gave some improvement in the calibration,
produced noticeably inferior results when the photometry on two images
of very different exposure times was compared. We had decided to stay
with the initial values for the correction coefficients
(Tab. \ref{tab:coeff}) and calibrate our photometry using a different
method. The coefficients for the color terms were derived from the
comparison of our NGC 6791 photometry with the data from the KPNO 0.9m
telescope (Kaluzny \& Udalski 1992). The offsets were determined
relative to the DIRECT catalog of stellar objects in M33 (Macri et
al. 2001b). The following transformations were adopted:
\begin{eqnarray*}
  v = V - 5.491 + 0.039\cdot(B-V)\\
b-v =  0.178 + 0.927\cdot(B-V)\\
  b = B - 5.312 - 0.034\cdot(B-V)
\end{eqnarray*}
The instrumental $V$ and $B$-band light curves of the variables were
transformed to the standard system by adding the appropriate offsets,
as the coefficients next to the color terms are very small.

Equatorial coordinates were determined for the $V$ template star list,
expanded with the variables with no $V$-band photometry. The
transformation from rectangular to equatorial coordinates was derived
using 818 transformation stars with $V<19.5$ from the DIRECT catalog
of stellar objects in M33 (Macri et al. 2001b). The average difference
between the catalog and the computed coordinates for the
transformation stars was $0.\arcsec07$ in right ascension and
$0.\arcsec06$ in declination. As a check we have also obtained an
independent transformation to the USNO-A2 catalog (Monet et al. 1996)
using 56 transformation stars ($\Delta\alpha=0.\arcsec26$,
$\Delta\delta=0.\arcsec23$). The average differences between the
computed coordinates and the M33 catalog for 1334 stars with $V<19.5$
were $\Delta\alpha=0.\arcsec71$ and $\Delta\delta=0.\arcsec60$.

\section{Catalog of Variables}
\subsection{Classification}

The variables we are most interested in are Cepheids and eclipsing
binaries (EBs). We therefore searched our sample of variable stars
primarily for these two classes of variables. The variable stars were
preliminarily classified as eclipsing, Cepheid or miscellaneous by
visual inspection, based on the shape of their light curves. The
variables for which neither $V$ nor $B$-band magnitude could be
determined (with only flux light curves or having a period in excess of
14 days) were not reclassified further. 

In order to obtain as clean a sample of Cepheids as possible, we have
inspected the location of the Cepheid variable candidates on a $V/B-V$
CMD. All of the Cepheid candidates having highly discrepant colors
were reclassified as other periodic variables. The candidates with
light curves in only one of the bands were checked on the
period-luminosity relation for that band. Extreme outliers were also
moved to the other periodic variable category.

The EBs with light curves expressed in magnitudes for at least one
band, were fitted with a model described in Papers I and II. Within
our assumption the light curve of an EB is determined by nine
parameters: the period, the zero point of the phase, the eccentricity,
the longitude of periastron, the radii of the two stars relative to
the binary separation, the inclination angle, the fraction of light
coming from the bigger star and the uneclipsed magnitude. Eclipsing
binary candidates for which a satisfactory fit was not achieved, were
reclassified as other periodic variables.

In the following sections \ref{sub:ecl}-\ref{sub:misc} we present the
parameters and light curves for the 434 identified variable
stars.\footnote{The $BV$ photometry and $V$ finding charts for all
variables are available from the authors via the {\tt anonymous ftp}
from the Harvard-Smithsonian Center for Astrophysics and can be also
accessed through the {\tt World Wide Web}.}  All tables are sorted by
increasing period, unless otherwise noted. We adopt a naming
convention after Macri et al. (2001a) based on the J2000.0 equatorial
coordinates, in the format: D33J$hhmmss.s$+$ddmmss.s$. The first three
fields ($hhmmss.s$) correspond to right ascension expressed in hours,
the last three ($ddmmss.s$) to declination, expressed in degrees,
separated by the declination sign. As none of the newly discovered
variables are present in previous variable star catalogs, we refer the
reader to Tables 5 and 6 in Paper VI for cross-identifications.

\subsection{Eclipsing Binaries}
\label{sub:ecl}

\begin{figure}[p]
\plotfiddle{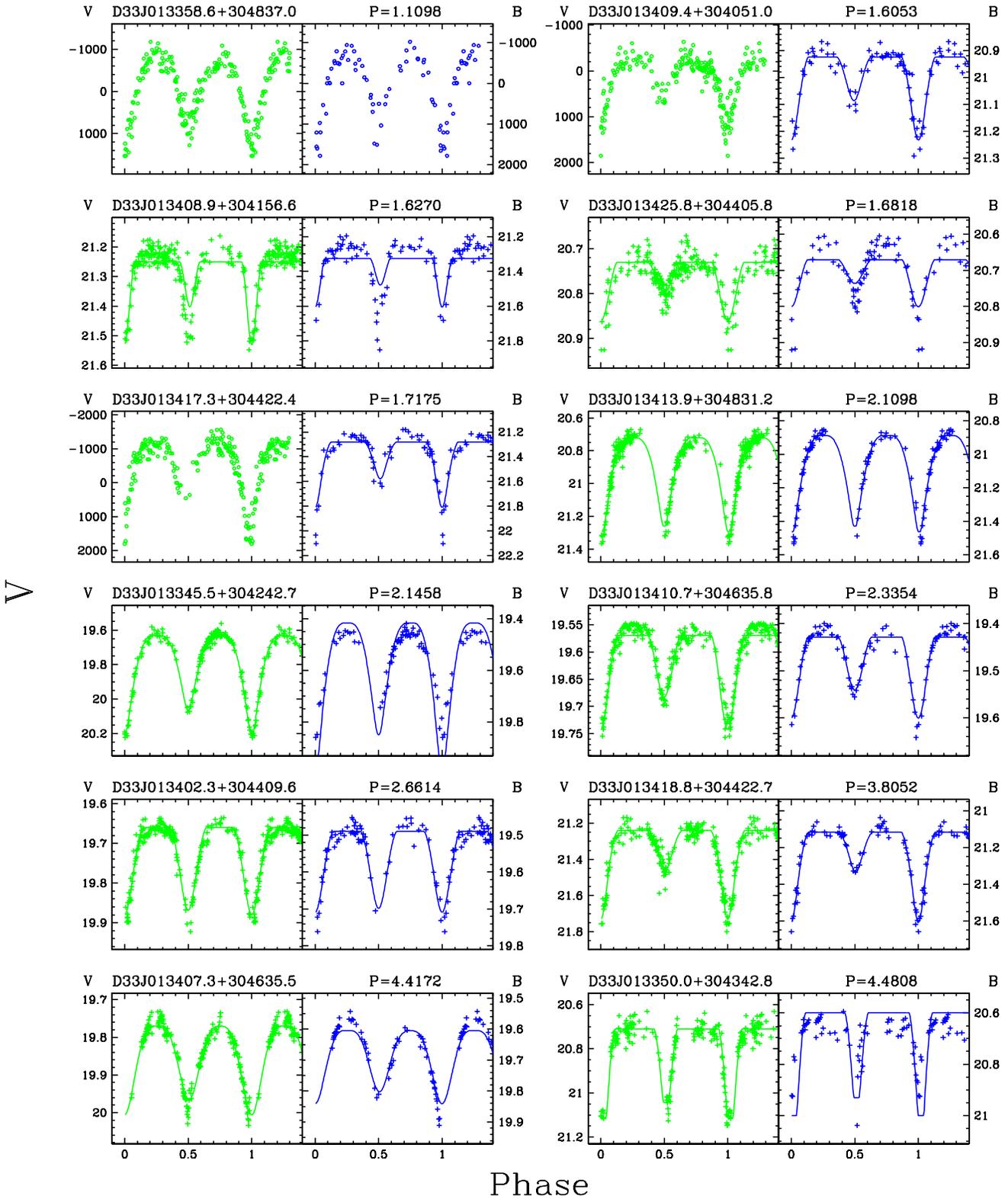}{19.5cm}{0}{88}{88}{-280}{-75}
\caption{Selected $BV$ light curves of eclipsing binaries found in the
field M33A. The points on the light curves expressed in magnitudes are
marked with crosses, the flux ones by open circles. The thin
continuous line represents the best fit model for each star and
photometric band.}
\label{fig:ecl}
\end{figure}

We have found a total of 63 eclipsing binaries in field M33A.  In
Table \ref{tab:ecl} we present the parameters for the 53 EBs with a
magnitude light curve in at least one band. For each variable we list
its name, period P, magnitudes $V_{max}$ and $B_{max}$ of the system
outside of the eclipse, and the radii of the binary components
$R_1,\;R_2$ in the units of the orbital separation.  We also give the
inclination angle of the binary orbit to the line of sight $i$ and the
eccentricity of the orbit $e$.  The reader should bear in mind that
the values of $V_{max},\;B_{max},\; R_1,\;R_2,\;i$ and $e$ are derived
with a straightforward model of the eclipsing system, so they should
be treated only as reasonable estimates of the ``true'' value. Table
\ref{tab:ecl_flux} lists 10 EBs with flux light curves only. For each
variable we give its name and period P. Figure \ref{fig:ecl} presents
the phased light curves of 12 sample EBs (see also Table 8). 

\subsection{Cepheids}
\label{sub:cep}

A total of 305 Cepheid variables were found in field M33A. In Table
\ref{tab:ceph} we present 242 Cepheids with a magnitude light curve in
at least one band. For each variable, we list its name, period P,
flux-weighted average magnitudes $\langle V\rangle$ and $\langle
B\rangle$ and the $V$ and $B$-band amplitudes $A_V$ and $A_B$. Due to
the short time base of our observations, reliable flux-weighted
magnitudes could only be determined for variables with periods shorter
than 14 days. We have extracted light curves for 36 Cepheids with
longer periods, all of them identified previously in Paper VI and made
them available via {\tt anonymous ftp}. In Table \ref{tab:ceph_flux}
we list 27 Cepheids with flux light curves only. For each variable we
list its name and period P. Figure \ref{fig:ceph} presents the phased
light curves of 24 sample Cepheids (see also
Table 9).

\begin{figure}[p]
\plotfiddle{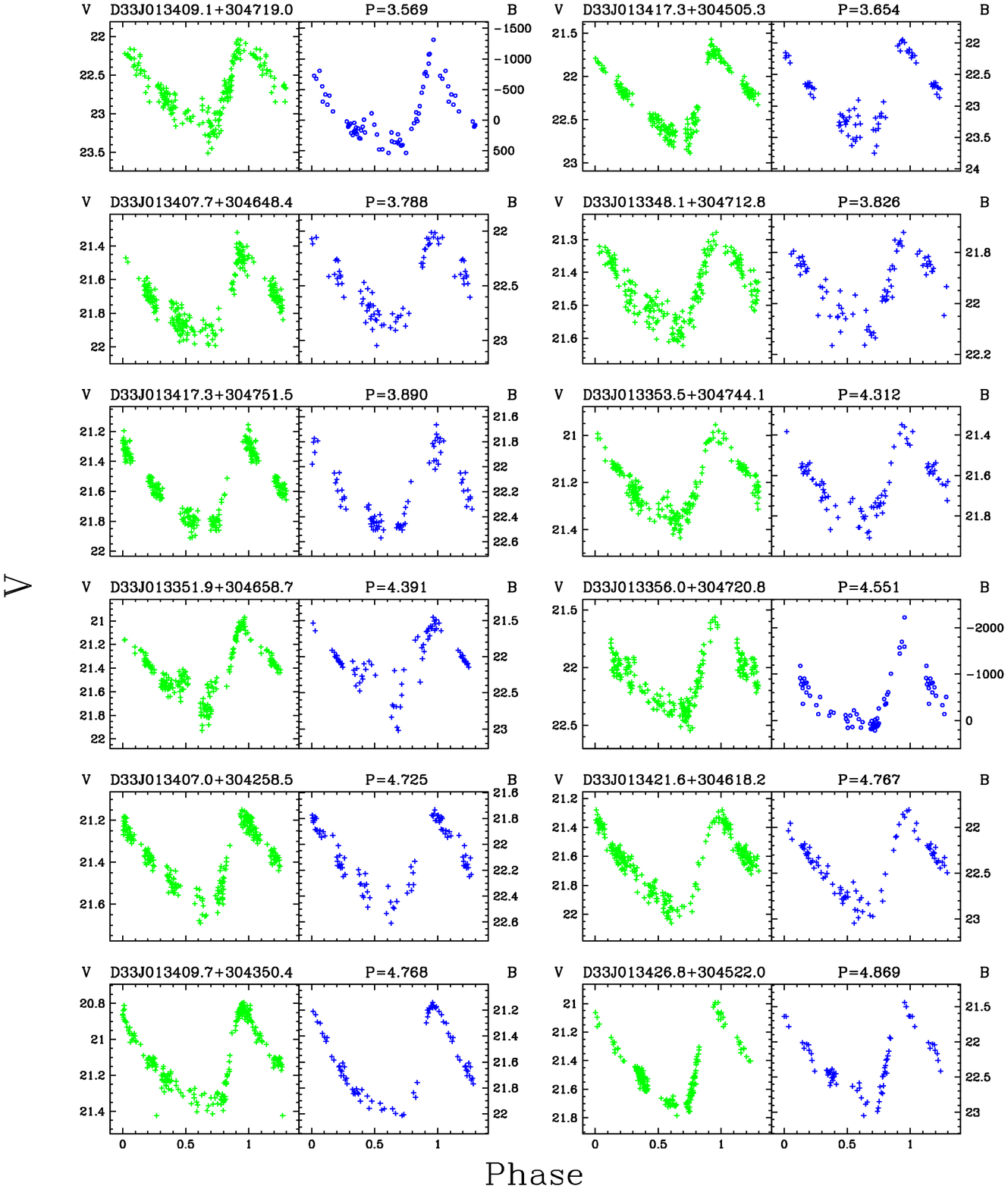}{19.5cm}{0}{88}{88}{-280}{-75}
\caption{Selected $BV$ light curves of Cepheid variables found in the
field M33A. The points on the light curves expressed in magnitudes are
marked with crosses, the flux ones by open circles.}
\label{fig:ceph}
\end{figure}

\addtocounter{figure}{-1}
\begin{figure}[p]
\plotfiddle{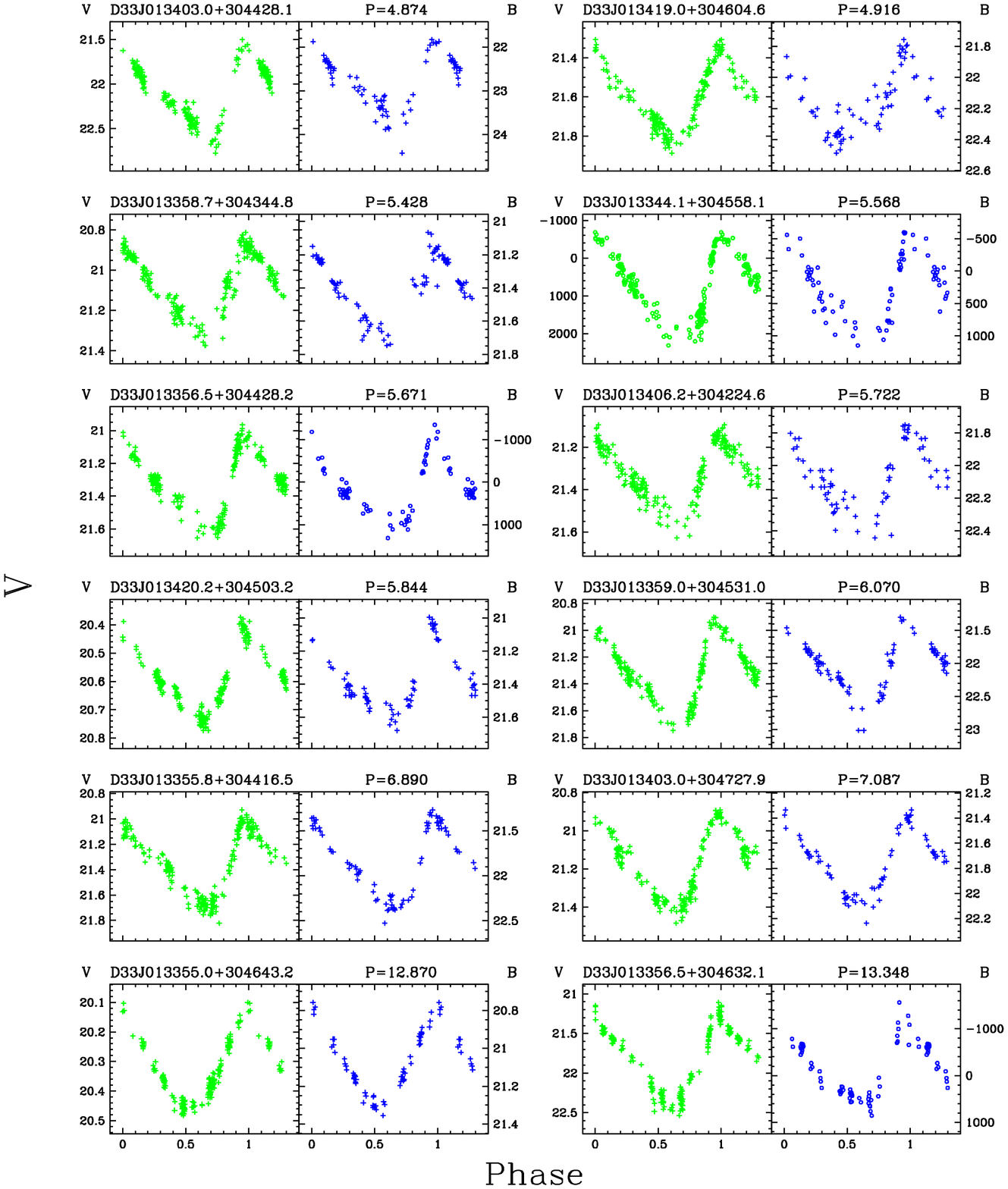}{19.5cm}{0}{88}{88}{-280}{-75}
\caption{Continued.}
\end{figure}

\begin{figure}[h]
\plotfiddle{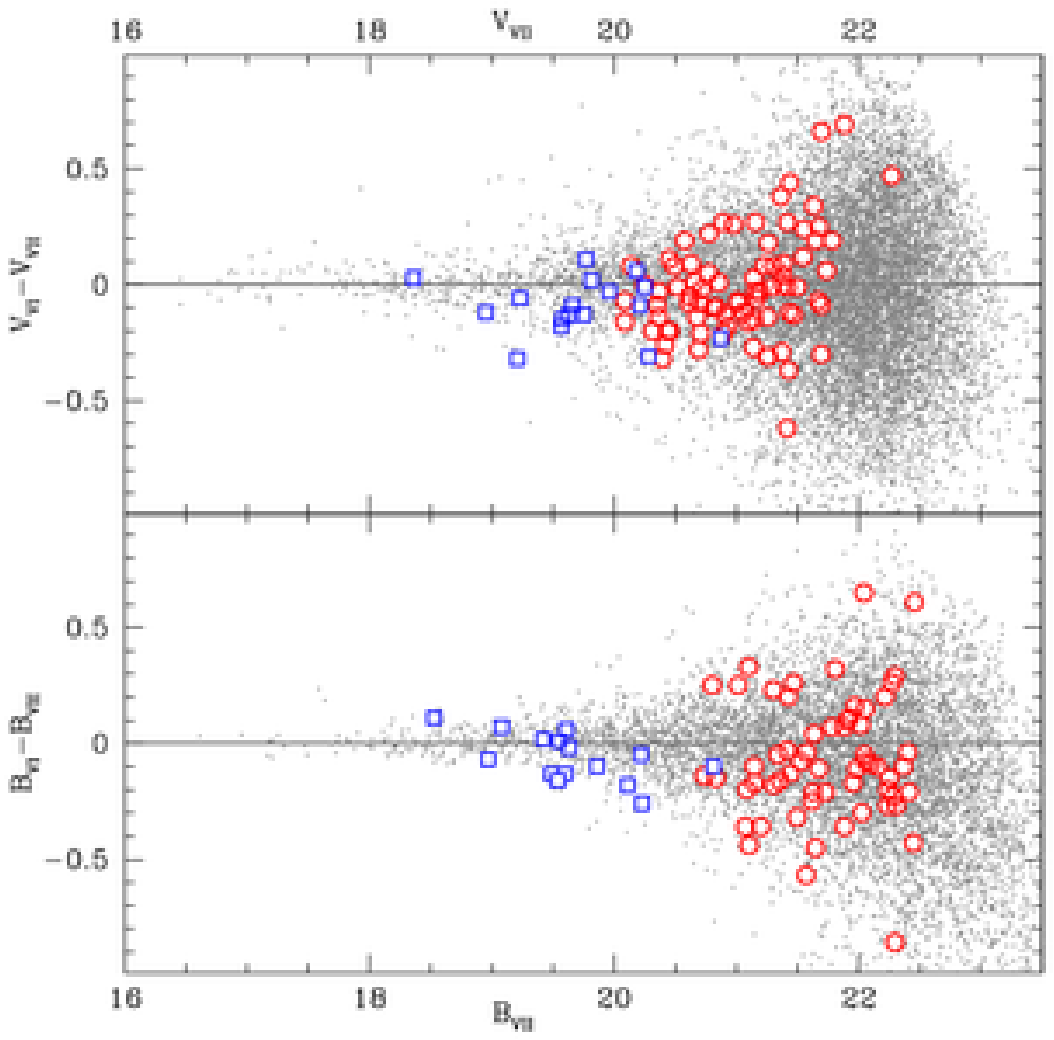}{9.7cm}{0}{100}{100}{-190}{-50}
\caption{A comparison between variable star photometry in the $V$
(upper panel) and $B$-band (lower panel) between our catalog and Paper
VI. EBs are denoted by squares and Cepheids by circles. All other
stars in the field are plotted with small dots in the background for
reference.}
\label{fig:cmp}
\end{figure}

\subsection{Other Periodic Variables}
\label{sub:per}
In Table \ref{tab:per} we present the parameters of 13 periodic
variables.  For each variable we list its name, period P, the
magnitudes $V$ and $B$ and the $V$ and $B$-band amplitudes $A_V$ and
$A_B$. The $V$ and $B$ columns list the magnitudes outside of the
eclipses $V_{max}$ and $B_{max}$ for the eclipsing variables and
flux-weighted average magnitudes $\langle V\rangle$ and $\langle
B\rangle$ for the other variables. We have also found three other
variables with periods of the order of 50 days (identified previously
in Paper VI), which we do not list in the table. Their light curves
are available via {\tt anonymous ftp}. In Table 10
we list the light curves of all periodic variables.

\subsection{Miscellaneous Variables}
\label{sub:misc}
In Table \ref{tab:misc} we present the parameters of 50 miscellaneous
variables.  For each variable we list its name, the average magnitudes
$\bar{V}$ and $\bar{B}$ and the $V$ and $B$-band amplitudes $A_V$ and
$A_B$. Variable D33J013342.3+304024.7 is probably an EB, with
incomplete phase coverage. Several of those variables are most likely
periodic, with periods in excess of 14 days. In
Table 11 we list the light curves of all miscellaneous
variables.

\begin{figure}[!h]
\plotfiddle{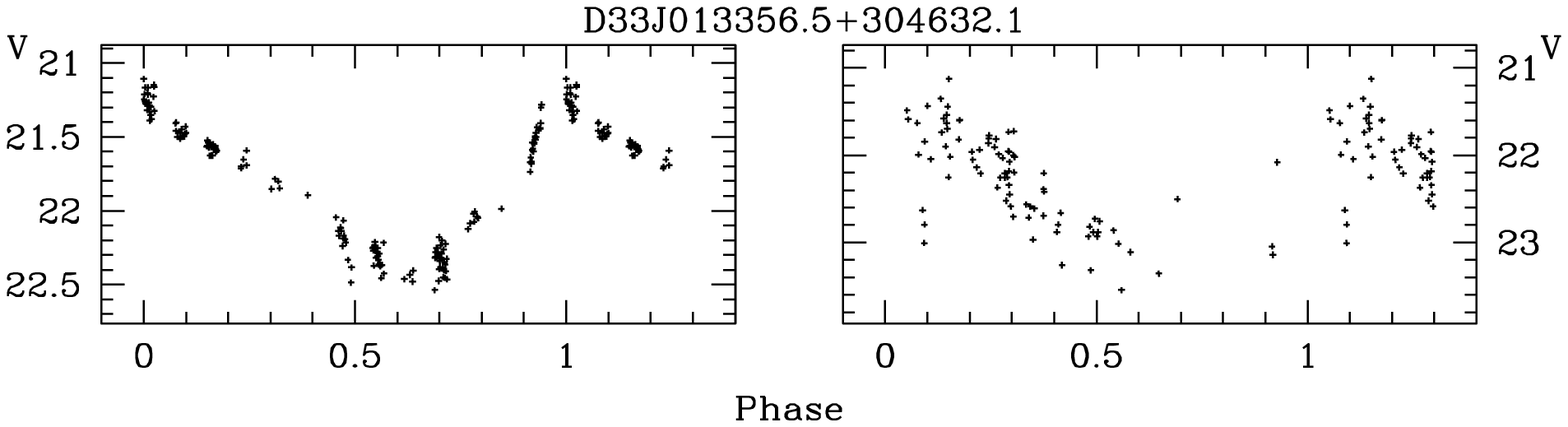}{3.7cm}{0}{85}{85}{-270}{-475}
\caption{A comparison of $V$-band light curves for the type II Cepheid
D33J013356.5+304632.1 (P=13.348 days) from our catalog and Paper
VI. The ISIS light curve obtained with the 2.1m telescope is shown in
the left panel and the {\sc daophot} light curve from 1.3m telescope
data in the right panel.}
\label{fig:cmp_lc}
\end{figure}

\subsection{Comparison with the catalog in Paper VI}
\label{sect:cmp}

In Figure \ref{fig:cmp} we compare our $V$ and/or $B$ photometry for
81 Cepheids (circles) and 18 EBs (squares) with the values listed in
the DIRECT catalog of variables in M33 (Paper VI). As reference we
plot in the background similar comparisons for all the stars in the
field (dots). The variables, with some exceptions, have $\Delta B$ and
$\Delta V$ distributions roughly similar to the rest of the stars,
although they do show a marked tendency to be fainter in our catalog.
This trend is especially prominent in the $V$-band for the EBs, but is
absent in their $B$-band comparison. We have examined the images of
three of the more discrepant EBs and found that in our data we have
resolved more nearby companions to these variables. These companions
are brighter in $V$ than in $B$, relative to the EBs. This is not
surprising, as the EBs are among of the bluest stars in the
field. There are also a few faint Cepheids, which show unusually large
differences in photometry, of the order of 0.6-0.8 mag. We have
inspected their light curves in both catalogs. The $V$-band light
curves of one such star, type II Cepheid D33J013356.5+304632.1, are
shown in Fig. \ref{fig:cmp_lc}. The left panel shows our light curve,
extracted with ISIS from 2.1m telescope data. The right panel shows
the {\sc daophot} light curve from data taken with a 1.3m telescope
(Paper VI). It seems that these discrepancies are caused in large part
by the fact that fixed position photometry is prone to identify and
fit a profile at the supplied position even when the star is below the
detection threshold, resulting in the false measurement of a fainter
magnitude.

\begin{figure}[h]
\plotfiddle{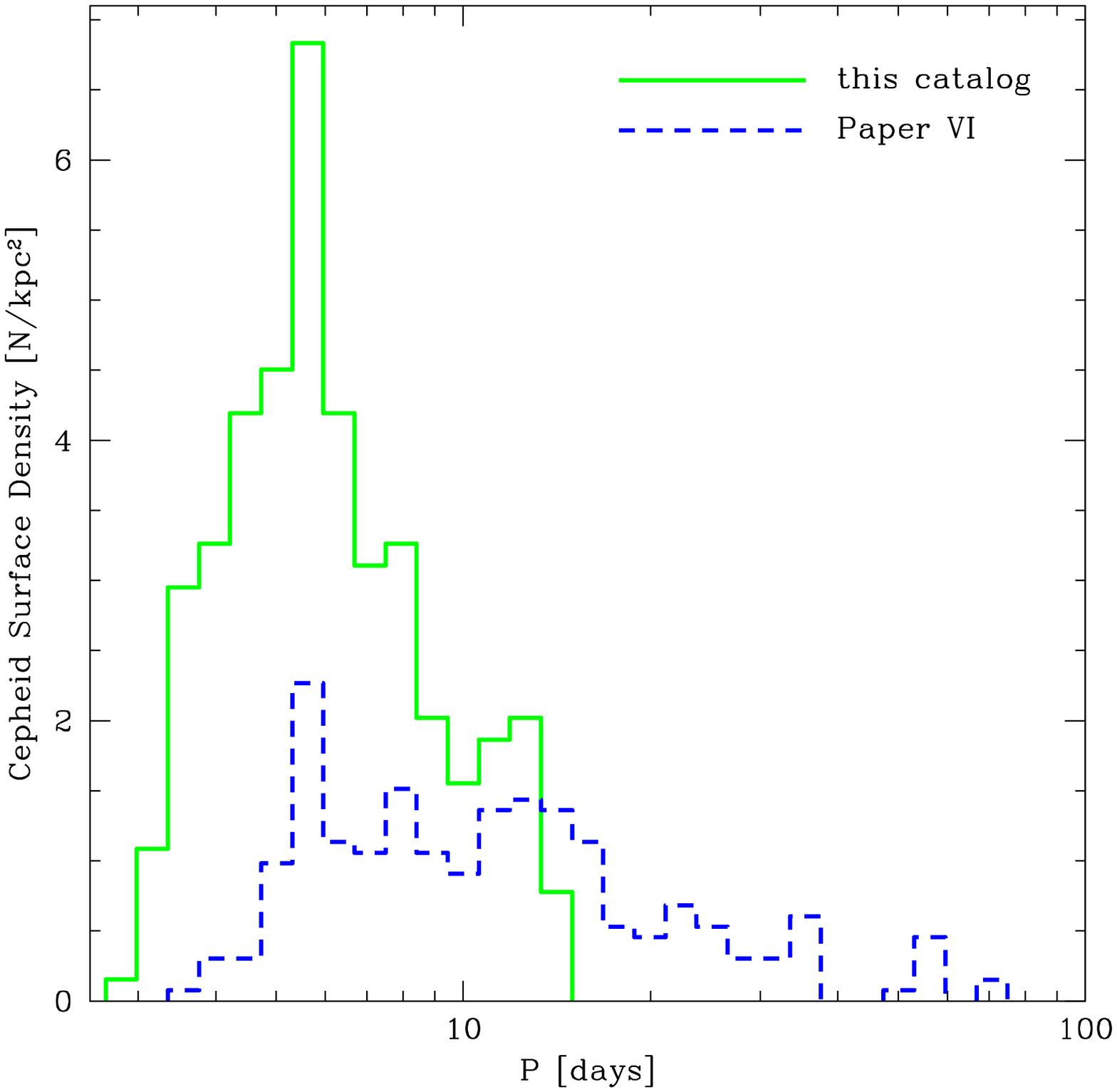}{9.7cm}{0}{55}{55}{-180}{-95}
\caption{A histogram of the surface density of Cepheids (N/arcmin$^2$)
as a function of their period. The solid line represents all Cepheids
from our catalog with $P<14$ days, the dashed line -- the Cepheids
from Paper VI.}
\label{fig:per}
\end{figure}

\begin{figure}[h]
\plotfiddle{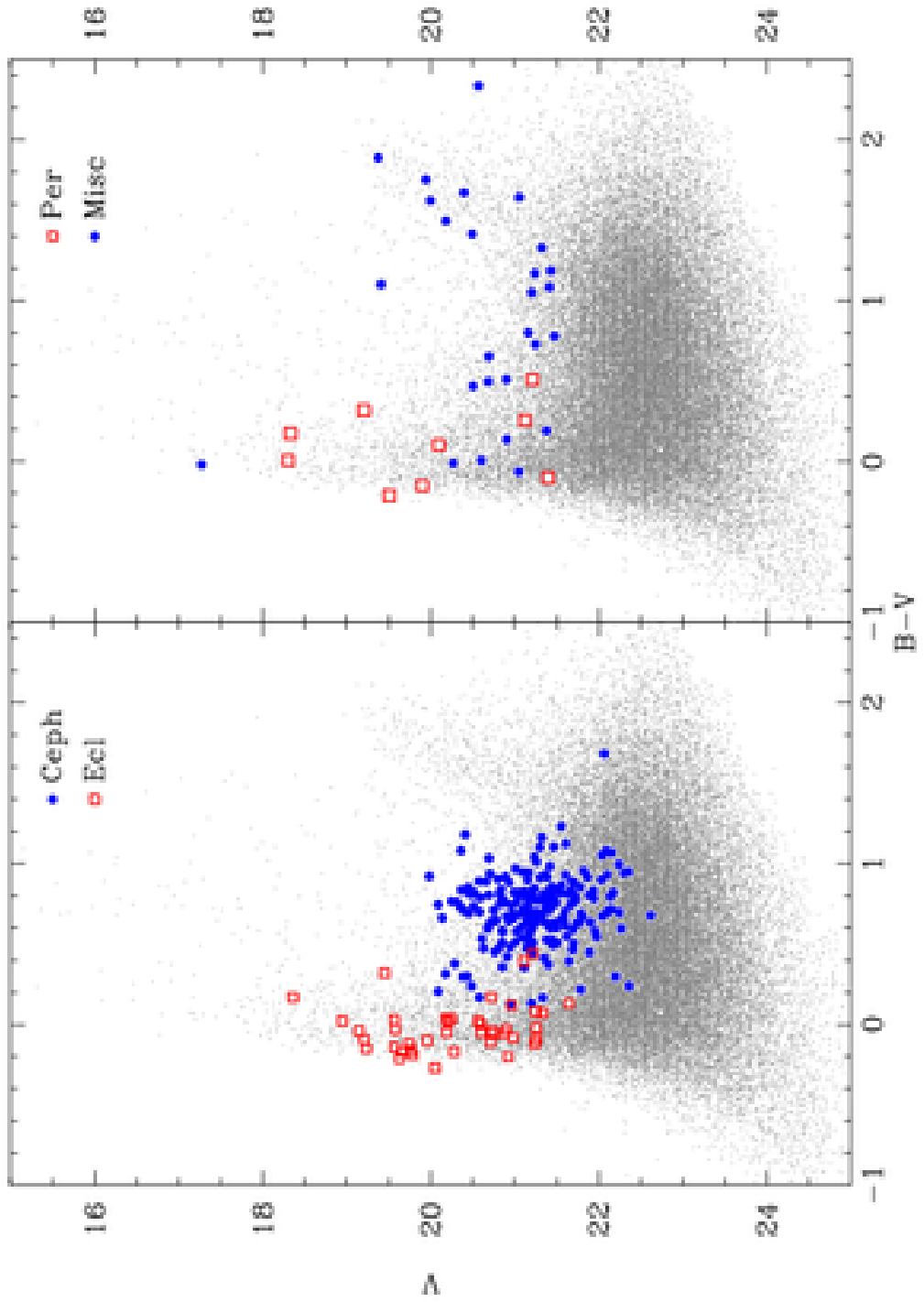}{9.0cm}{-90}{100}{100}{-250}{330}
\caption{The $V/B-V$ CMD for the variable stars in field M33A. The
EBs (open squares) and Cepheids (filled circles) are shown in 
the left panel, periodic (filled squares) and miscellaneous 
variables (open circles) in the right panel.}
\label{fig:cmd}
\end{figure}

In Figure \ref{fig:per} we plot a histogram of the surface density of
Cepheids (N/kpc$^2$) as a function of their period, assuming a
distance of 840 kpc to M33 (Freedman, Wilson \& Madore 1991). The
solid line represents all Cepheids from our catalog with $P<14$ days,
the dashed line -- the Cepheids from Paper VI. The areas covered in
this search and in Paper VI are 108 and 222 arcmin$^2$. As our
observations were carried out during two one-week runs spaced one
month apart, we lack the baseline to detect long period
Cepheids. Since our data were collected with an instrument 2.6 times
larger in area and with better seeing, we have a higher detection rate
for short period Cepheids. The Paper VI catalog, due to the much
longer baseline of observations, contains more long period Cepheids.

\section{Discussion}

In Fig. \ref{fig:cmd} we plot the positions of the variable stars on
the $V/B-V$ CMD. The EBs are denoted by open squares and Cepheids 
by filled circles in the left panel, the periodic variables by 
filled squares and miscellaneous by open circles in the right
panel.

All of the EBs occupy the upper main sequence, with the exception of
D33J013359.5+304037.6 ($B-V=0.37$). This variable most likely suffers
from larger than average reddening, as it is located at the edge of 
what appears to be a dust lane. 

\begin{figure}[h]
\plotfiddle{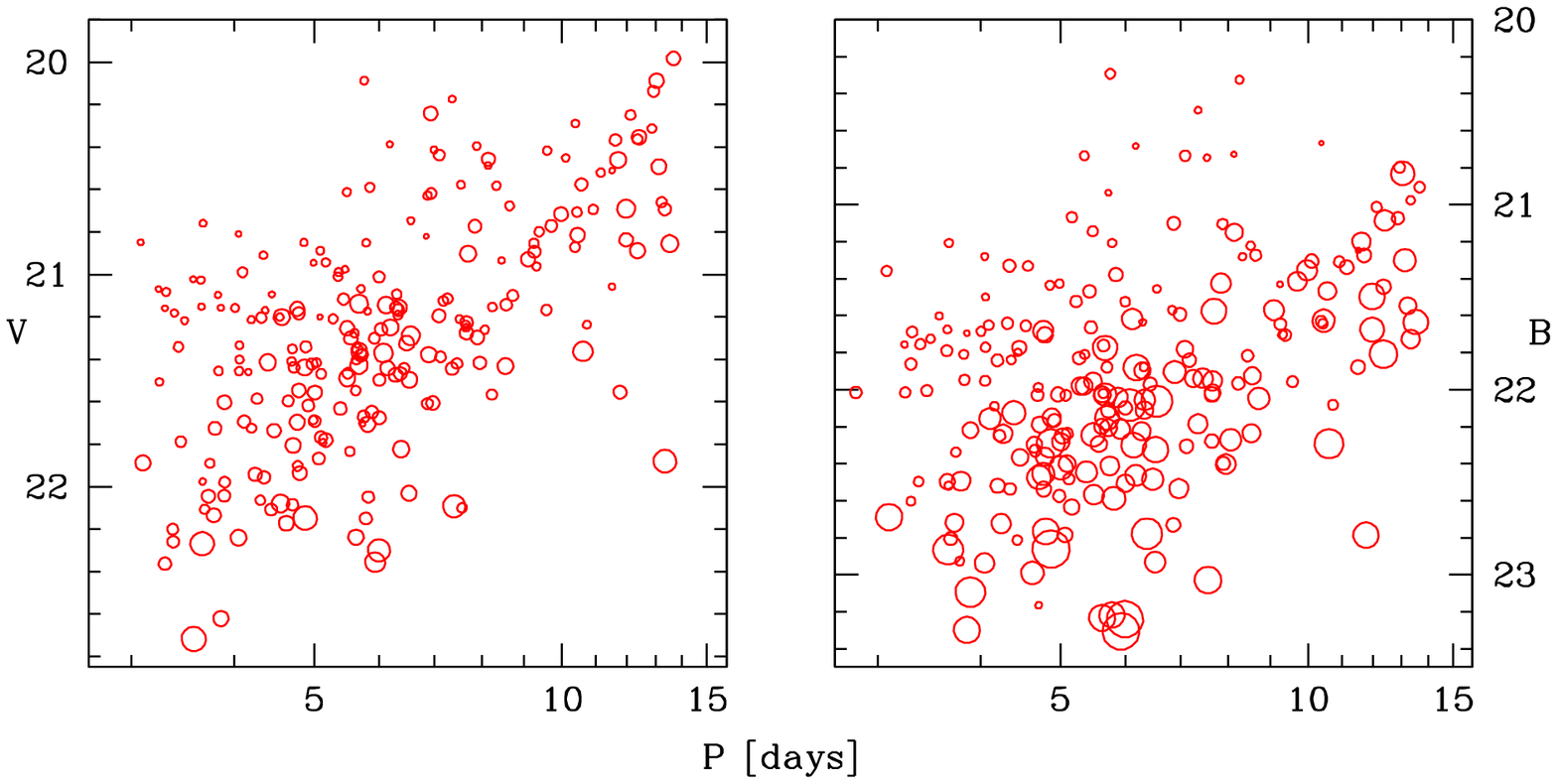}{8cm}{0}{90}{90}{-280}{-385}
\caption{The $V$ and $B$-band Period-Luminosity diagrams for the 
Cepheids in field M33A. The circles are proportional in size to 
their amplitudes in the corresponding band.}
\label{fig:pl}
\end{figure}

Most Cepheids occupy the region between $0.3<B-V<1.0$, with several
outliers stretching from $B-V=0.10$ to $B-V=1.24$. The discrepant
colors of some of the Cepheid variables most likely originate from the
phenomenon of blending with nearby bright stars of very different
colors and/or reddening. The phenomenon of blending occurs when the
Cepheid possesses one or more close companions which cannot be
separated at the resolving power of the instrument used. A further
discussion of blending and its properties can be found in Mochejska et
al. (2000; 2001) and Stanek \& Udalski (1999).

All but one of the periodic variables with $BV$ photometry are bluer
than $B-V=0.35$. The five brightest ones exhibit Cepheid-like light
curves of very low amplitude of the order of a few hundredths of a
magnitude. The low amplitudes of their variability suggest that these
could be Cepheids blended with blue main sequence stars. The remaining
four periodic variables are likely EBs. Three of them have blue
colors, consistent with this type of variable. The fourth one,
D33J013346.1+304658.9, with a period of 6.4640 days seems to be a
contact EB, despite its unusual color ($B-V=0.511$). A more complete
light curve would be required to resolve the true nature of this
object.

The miscellaneous variables are spread throughout the CMD. Some are
located on the upper main sequence, and a few seem to fall on the
upper red giant branch. Several seem to occupy the same region as
Cepheids. It is possible that some of these variables are Cepheids
with periods in excess of 14 days.

\begin{figure}[!ht]
\plotfiddle{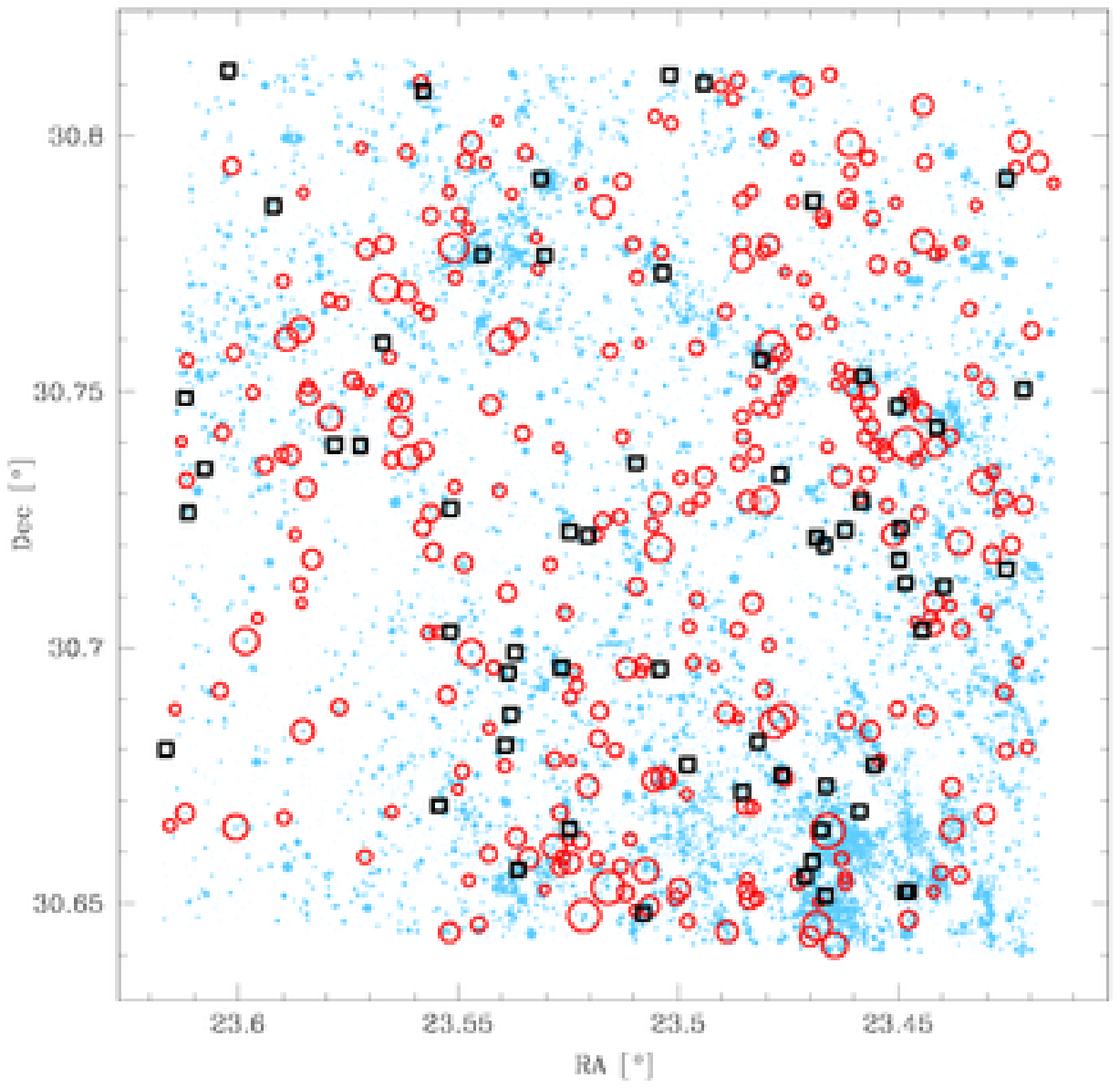}{12cm}{0}{100}{100}{-250}{-50}
\caption{The location of EBs (squares) and Cepheids (circles) in 
the field of M33A. The size of the Cepheid symbols are proportional
to their period. All other stars in the field with $V<21.5$ mag are
drawn as dots of size proportional to their magnitude.}
\label{fig:xy}
\end{figure}

In Figure \ref{fig:pl} we present the $B$ and $V$-band P-L diagrams
for the Cepheid variables, drawn as open circles proportional in size
to their amplitudes. As expected, the amplitudes in the $B$-band are
on average larger than in $V$. A clear relation between the period and
magnitude is discernible. The several Cepheids too faint for their
periods are probably suffering from greater than average reddening.
Some of the faintest Cepheids exhibit quite large amplitudes. This is
not a physical effect and is most likely caused by a similar
phenomenon to the one discussed in Subsection \ref{sect:cmp} regarding
the comparison with Paper VI photometry. If the magnitude used to
convert the light curve from flux to magnitudes is measured too faint,
the resulting amplitude will be too large.

For periods shorter than 8 days the relations widen upwards
considerably, with the brighter Cepheids for a given period having
smaller amplitudes. One possibility is that these Cepheids are
pulsating in the first overtone. On the other hand these could be
fundamental mode Cepheids which are heavily affected by blending:
adding a constant flux would tend to increase the brightness of a
Cepheid and diminish its amplitude. We will examine these stars in
more depth in Section \ref{sect:fo}.

\begin{figure}[h]
\plotfiddle{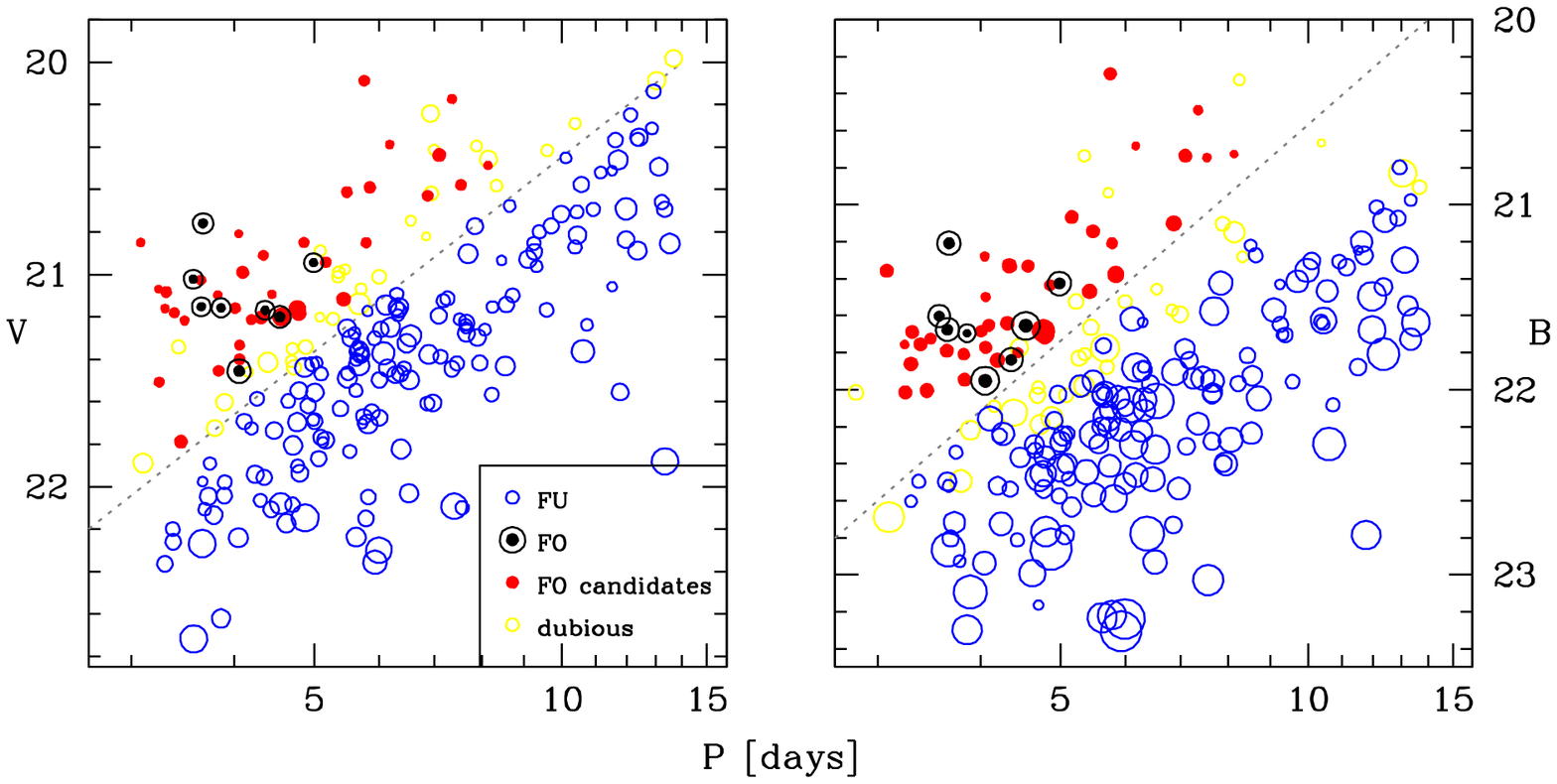}{8cm}{0}{90}{90}{-280}{-385}
\caption{The $V$ and $B$-band Period-Luminosity diagrams for the
Cepheids in field M33A. The circles are proportional in size to the
pulsational amplitudes of the Cepheids in the corresponding band. The
dotted lines show the division between FU (open circles) and FO
(filled circles) Cepheids. The encircled dots represent the most
reliable FO Cepheid candidates. The light open circles show the
Cepheids which are above the line in one band only.}
\label{fig:pl_fo}
\end{figure}

Figure \ref{fig:xy} shows the location of the EBs and Cepheids within
the field M33A. The Cepheids are plotted as open circles, proportional
in size to their period and the EBs as open squares. To trace the
spiral pattern of the galaxy we plot in the background all the stars
with $V<21.5$ mag as filled dots of size proportional to their
magnitude. These two types of variables appear somewhat more plentiful
within the spiral arms.

\section{First Overtone Cepheids}
\label{sect:fo}

As we have noted in the previous Section, on the $B$ and $V$-band P-L
diagrams in Fig. \ref{fig:pl} there are Cepheids which seem too bright
for their periods. In addition they possess smaller pulsational
amplitudes of variability compared to the normal Cepheids. Their
positions on the P-L diagrams would lead us to expect that these
should be first overtone (FO) pulsators (as in Fig. 2 of Udalski et
al. 1999; hereafter U99). The situation is, however, complicated by
the existence of blending. As a result of blending the Cepheid should
appear brighter because of the added constant flux and its amplitude,
measured in magnitudes, should decrease (Mochejska et al. 2000,
2001; Stanek \& Udalski 1999).

\begin{figure}[h]
\plotfiddle{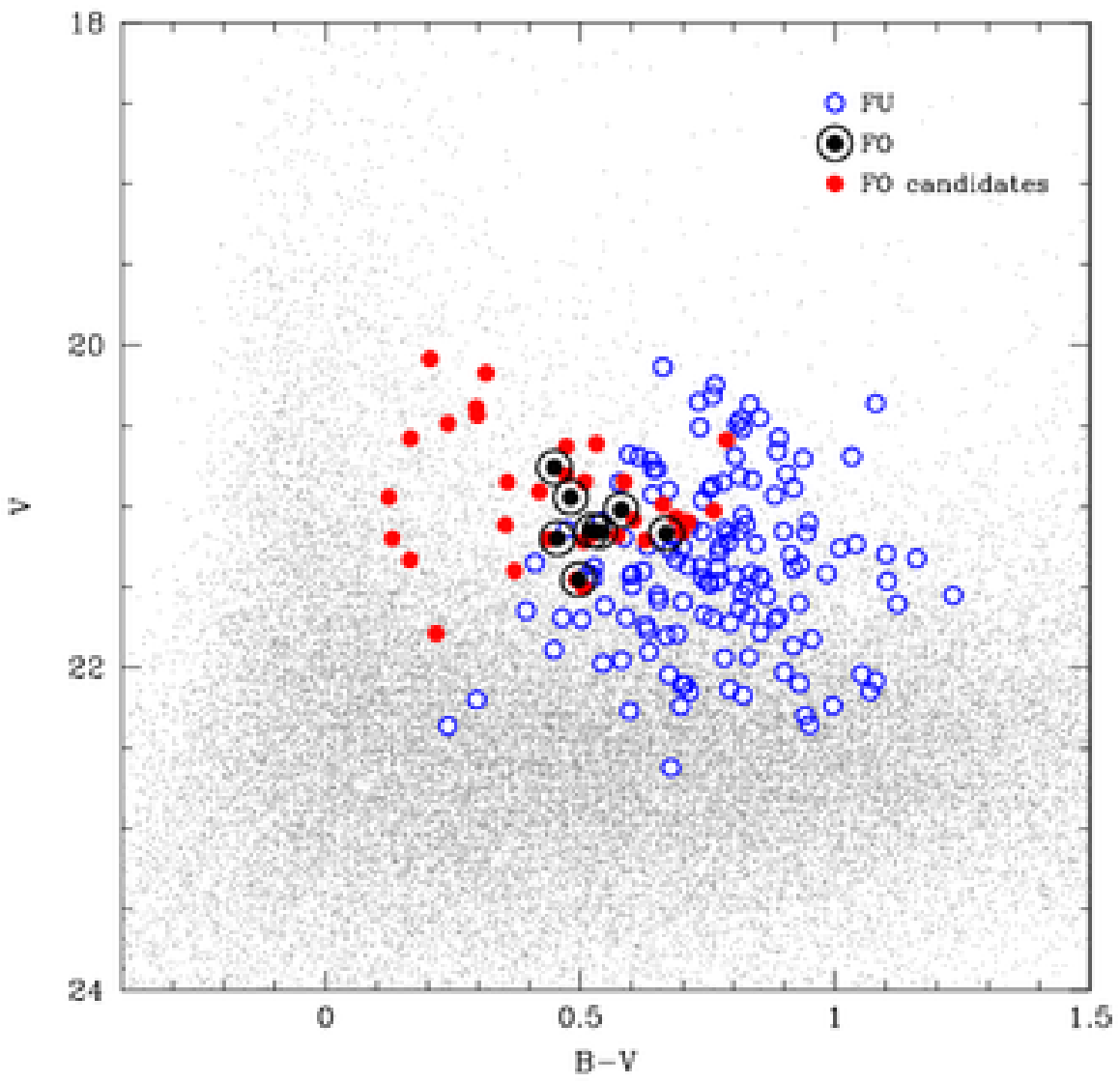}{10.5cm}{0}{100}{100}{-230}{-50}
\caption{The $V/B-V$ CMD for the Cepheid variables. The FU Cepheids are
denoted by open circles and the FO ones by filled circles. The encircled
dots represent the most reliable FO Cepheid candidates.}
\label{fig:cmd_fo}
\end{figure}

In order to try to determine whether these Cepheids are first overtone
pulsators, we have checked whether they possess other properties
expected of such stars. In addition to being brighter and having a
smaller amplitude, Cepheids pulsating in the first overtone should
have more symmetric (sinusoidal) light curves and be on average bluer
than fundamental mode Cepheids (FU). The most powerful technique for
discriminating them from FU Cepheids are the Fourier parameters of
their light curves (Antonello \& Aikawa 1995; Beaulieu et al. 1995).

To select a sample of FO Cepheid candidates we have made a division
on the P-L diagrams roughly parallel to the P-L relation, between the
bright low amplitude and fainter high amplitude Cepheids (dotted lines
on Fig. \ref{fig:pl_fo}). In our final sample we included the Cepheids
which were above these lines in both of the P-L diagrams. There are 44
such Cepheids in our catalog (filled circles). FO Cepheids should have
periods ranging from 1.7 to 6 days.  We decided not to make a cutoff
at higher periods, although we did not regard it likely that Cepheids
with $P>6$ days would turn out to be FO pulsators.

\begin{figure}[h]
\plotfiddle{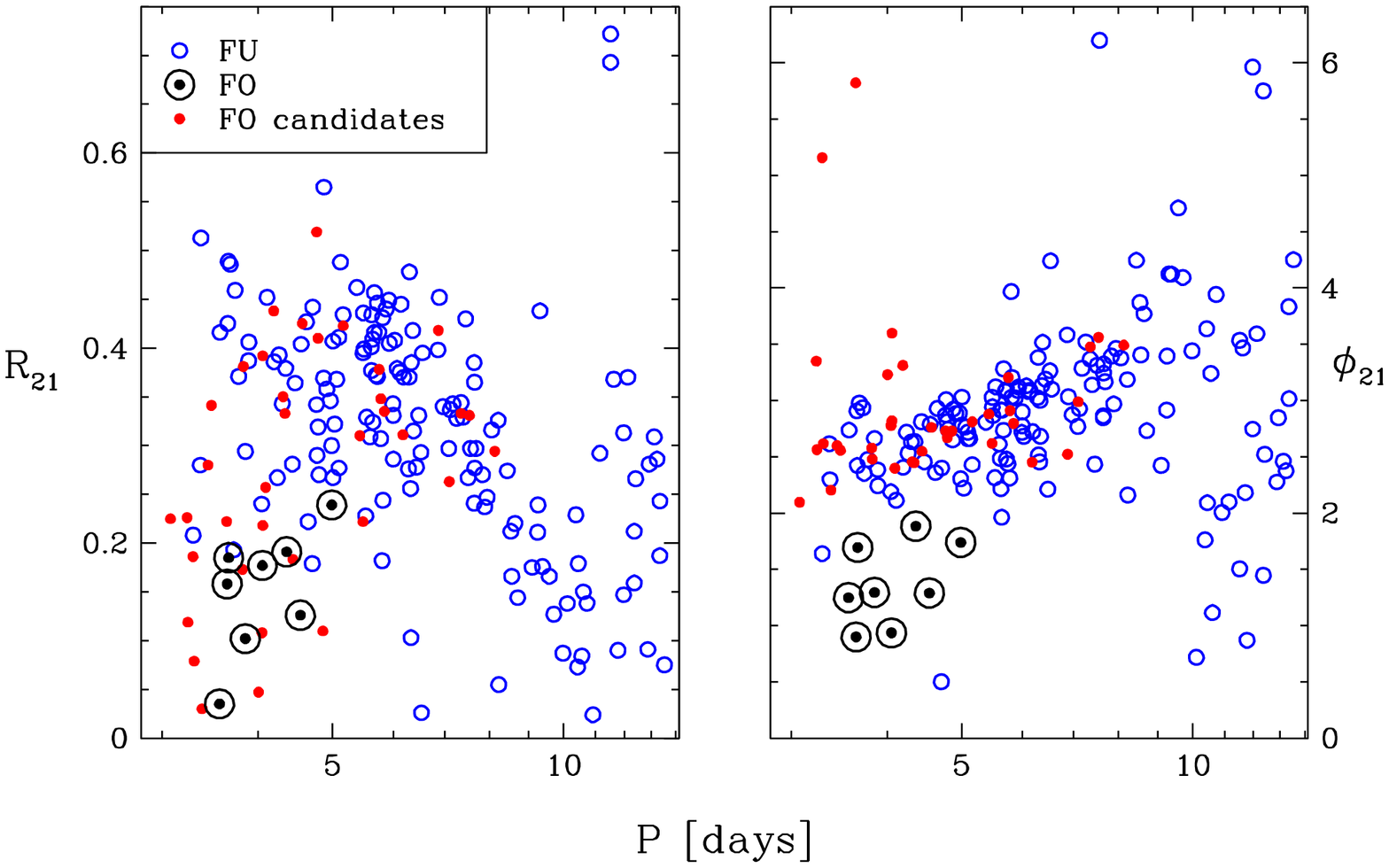}{9.2cm}{0}{80}{80}{-260}{-315}
\caption{The Fourier parameters $R_{21}$ and $\phi_{21}$ of the
Cepheid light curves as a function of period. The FU Cepheids are
denoted by open circles and the FO ones by filled circles. The
encircled dots represent the most reliable FO Cepheid candidates.}
\label{fig:pars}
\end{figure}

On the $V/B-V$ CMD (Fig. \ref{fig:cmd_fo}) we plot the positions of
the FU Cepheids (open circles) and FO candidates (filled circles). As
expected of FO Cepheids, our candidates are bluer than the FU ones. We
note that on the CMD presented by U99 for the Large Magellanic Cloud
(LMC) Cepheids (Fig. 4 therein) there is also some overlap between the
loci of those two types of Cepheids.

The relations between the light curve Fourier parameters
$R_{21}=A_2/A_1$, $\phi_{21}=\phi_2-2\phi_1$ and the period for the FO
Cepheid candidates (filled circles) and FU mode pulsators (open
circles) are shown in Fig. \ref{fig:pars}. We have compared them with
Fig. 3 in U99 for LMC Cepheids. We notice on the $R_{21}/\log P$
diagram for LMC that within our range of periods the FO Cepheids
progress upwards in $R_{21}$ with increasing period, forming the
second branch of the V-shaped pattern and then merge with the FU
Cepheid sequence at higher periods. The situation is very similar on
the $\phi_{21}/\log P$ diagram in U99, where on the one hand the FU
Cepheids are confined to a narrower sequence, but on the other there
is more overlap between them and the FO pulsators.

\begin{figure}[h]
\plotfiddle{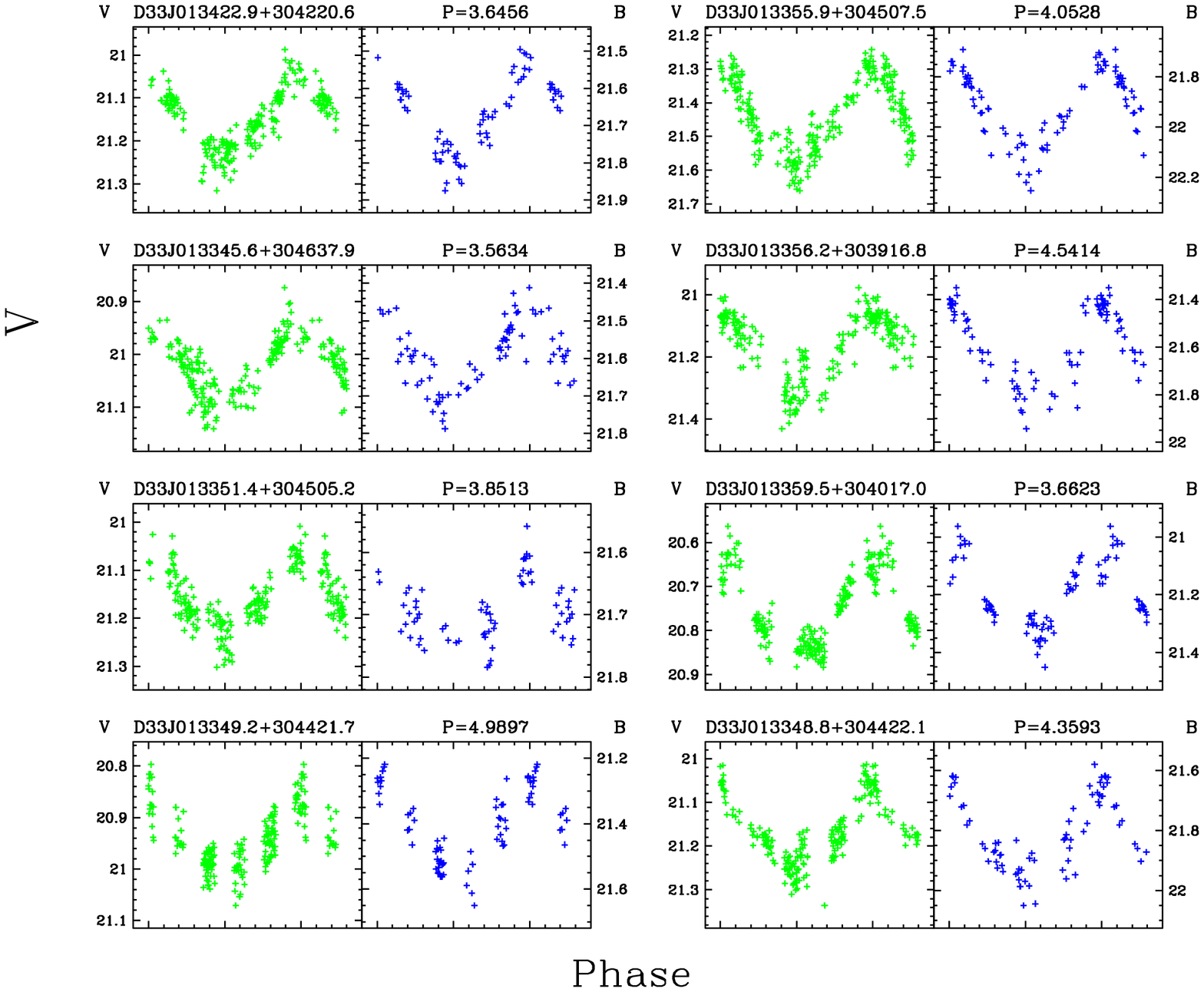}{13cm}{0}{88}{88}{-280}{-260}
\caption{The $V$ and $B$-band light curves of the eight FO Cepheids
in the field M33A.}
\label{fig:lc_fo}
\end{figure}

Indeed we do observe in Fig. \ref{fig:pars} the FO Cepheid candidates
to occupy roughly the predicted positions despite our larger scatter
in the Fourier parameters than in U99. From the $\phi_{21}/\log P$
diagram in the right panel we have selected eight FO Cepheids which
are most separated from the FU Cepheid sequence. We examined their
positions on the $R_{21}$ diagram to find that they are also rather
well separated from the FU mode Cepheids, especially the seven with
shorter periods. These eight bona fide FO Cepheid candidates are
denoted by encircled dots on Figs \ref{fig:pl_fo}-\ref{fig:pars}. 

We have checked the brightest of these Cepheids,
D33J013359.5+304017.0, which is unusually bright even for an overtone
Cepheid, in archival HST data. It has two nearby companions, one of
which appears bright enough to affect its photometry. To try to
estimate the influence of blending on the Fourier parameters $R_{21}$
and $\phi_{21}$ we have generated two new light curves for each of the
eight best FO Cepheid candidates, one with an $S_V =
\frac{f_blend}{f_ceph} = 0.5$ blend added and one with the same blend
subtracted. The average differences between the original and
recomputed parameters were 9\% for $R_{21}$ and 6\% for
$\phi_{21}$. These differences are too small to change the positions
of the Cepheids considerably on the $R_{21}$ and $\phi_{21}$ diagrams.
This very weak dependence of the Fourier parameters on blending
strengthens our conclusion that these Cepheids are indeed first
overtone pulsators.

As for the rest of the candidates, we believe many of them may also be
FO Cepheids. More accurate photometry obtained with a larger
instrument would be necessary to obtain better Fourier parameters of
their light curves and resolve this issue.

In Figure \ref{fig:lc_fo} we show the light curves of the eight FO
Cepheids. They appear more symmetrical in shape and do not exhibit the
fast rise and slow decline typical for FU Cepheids (see Fig.
\ref{fig:ceph}). This fact lends further credibility to the notion 
that these Cepheids pulsate in the first overtone.

\section{Conclusions}
Our search for variable stars in the data from the followup
observations of the detached eclipsing binary D33J013346.2+304439.9 in
field M33A collected at the 2.1m KPNO telescope resulted in the
discovery of 63 eclipsing binaries, 305 Cepheids, and 66 other
periodic, possible long period or non-periodic variables. Out of the
total 434 variables 280 are not listed in our first M33 catalog (Paper
VI). Due to the short time base of our observations, we were limited
to variables with periods not exceeding 14 days. Thanks to the use of
a larger aperture instrument and a novel method of image reduction --
the ISIS image subtraction package, we were more efficient at finding
the fainter and lower amplitude variables than in our previous study
of this field, especially for short period Cepheids ($P<8d$). We have
also found a population of Cepheids which are most likely pulsating in
the first overtone and for eight of them we present strong arguments
in favor of this interpretation.

The method of image subtraction has two main advantages over the
classical profile fitting method. It is more efficient in discovering
variables: we have discovered 355 periodic variables using ISIS and
only 212 with Dophot. Additionally in crowded fields image subtraction
can lead to large improvements in the photometric accuracy (Alard
2000b).

During our first observing run we have found that the T2KA CCD camera
produced a non-linear response to signal. We have obtained a set of
calibration frames to determine a correction for the non-linearity to
be applied to the science data. Although the comparison between images
of differing exposure times (Fig. \ref{fig:lincmp}) and with the Macri
et al. (2001b) stellar object catalog (Fig. \ref{fig:cmp}) do not show
marked non-linearity, some traces of it still remain. The presence of
a residual non-linearity prevented us from obtaining an independent
calibration. During the analysis of the data collected at KPNO in
October 1998 with the T1KA camera we have found it to be also
non-linear. We would like to caution the astronomical community of the
problems we have encountered with the T1KA and T2KA cameras.

\acknowledgments{We thank the TAC of the KPNO for the generous
allocation of the observing time. We would like to thank Lucas Macri
for his help and for the analysis of T1KA non-linearity, Grzegorz
Pojma{\'n}ski for $lc$ - the light curve analysis utility and Wojtek
Pych for his software. We thank the referee, Douglas Welch, for a
prompt and useful report. BJM and JK were supported by the Polish KBN
grant 2P03D003.17. BJM was also supported by the Polish KBN grant
2P03D025.19 and JK by the NSF grant AST-9819787. Support for KZS was
provided by NASA through Hubble Fellowship grant HF-01124.01-A from
the Space Telescope Science Institute, which is operated by the
Association of Universities for Research in Astronomy, Inc., under
NASA contract NAS5-26555.  DDS acknowledges support from the Alfred
P. Sloan Foundation and from NSF grant No. AST-9970812.}

\begin{small}
\tablenum{2}
\begin{planotable}{lrrrrrcrl}
\tablewidth{40pc}
\tablecaption{\sc DIRECT Eclipsing Binaries in M33A}
\tablehead{
\colhead{Name} & \colhead{$P$} & \colhead{} & \colhead{} & \colhead{} &
\colhead{} & \colhead{$i$} & \colhead{} & \colhead{} \\
\colhead{} & \colhead{$(days)$} & \colhead{$V_{max}$} & \colhead{$B_{max}$} &
\colhead{$R_1$} & \colhead{$R_2$} & \colhead{(deg)} & \colhead{$e$} &
\colhead{Comments}}
\startdata
 D33J013413.0+304008.8 &  0.9574 &   20.90 &   20.98 & 0.58 & 0.37 & 51.94 & 0.03 &     \\ 
 D33J013422.1+304710.4 &  1.0530 &   19.86 &   19.96 & 0.66 & 0.34 & 61.62 & 0.00 & (1) \\ 
 D33J013424.5+304845.8 &  1.0699 & \nodata &   21.03 & 0.62 & 0.35 & 58.18 & 0.00 &     \\ 
 D33J013400.4+304842.7 &  1.2517 &   21.24 &   21.26 & 0.39 & 0.26 & 71.20 & 0.01 &     \\ 
 D33J013352.1+303951.4 &  1.2529 &   21.18 & \nodata & 0.42 & 0.26 & 73.13 & 0.01 &     \\ 
 D33J013347.6+303908.3 &  1.3211 &   20.23 & \nodata & 0.39 & 0.29 & 59.22 & 0.00 &     \\ 
 D33J013427.9+304048.3 &  1.3919 & \nodata &   20.70 & 0.52 & 0.47 & 57.87 & 0.03 &     \\ 
 D33J013426.9+304455.5 &  1.4173 &   20.87 &   20.90 & 0.48 & 0.28 & 62.79 & 0.01 &     \\ 
 D33J013352.4+304317.2 &  1.4869 &   20.45 & \nodata & 0.63 & 0.36 & 58.91 & 0.00 &     \\ 
 D33J013352.5+304317.0 &  1.5229 &   20.45 & \nodata & 0.56 & 0.38 & 58.74 & 0.00 &     \\ 
 D33J013350.9+304322.6 &  1.5906 &   20.14 &   20.19 & 0.51 & 0.36 & 58.19 & 0.00 &     \\ 
 D33J013409.4+304051.0 &  1.6053 &   20.93 & \nodata & 0.46 & 0.33 & 65.80 & 0.01 &     \\ 
 D33J013408.9+304156.6 &  1.6270 &   21.33 &   21.25 & 0.31 & 0.21 & 75.93 & 0.02 &     \\ 
 D33J013425.8+304405.8 &  1.6818 &   20.67 &   20.73 & 0.43 & 0.40 & 53.71 & 0.00 &     \\ 
 D33J013417.3+304422.4 &  1.7175 &   21.28 & \nodata & 0.42 & 0.30 & 82.52 & 0.02 &     \\ 
 D33J013407.5+304729.8 &  1.8102 &   19.11 &   19.15 & 0.62 & 0.35 & 34.34 & 0.03 &     \\ 
 D33J013354.4+304401.7 &  1.8483 &   20.81 &   20.87 & 0.62 & 0.38 & 57.05 & 0.00 & (1) \\ 
 D33J013347.4+303908.0 &  1.8775 &   20.59 &   20.57 & 0.52 & 0.36 & 68.61 & 0.03 &     \\ 
 D33J013413.9+304831.2 &  2.1098 &   20.89 &   20.72 & 0.54 & 0.44 & 84.89 & 0.01 &     \\ 
 D33J013345.5+304242.7 &  2.1458 &   19.42 &   19.63 & 0.55 & 0.43 & 82.62 & 0.00 & (1) \\ 
 D33J013359.5+304037.6 &  2.1554 &   21.66 &   21.22 & 0.66 & 0.34 & 80.76 & 0.01 &     \\ 
 D33J013347.9+304323.8 &  2.2497 &   21.31 & \nodata & 0.48 & 0.31 & 54.45 & 0.09 &     \\ 
 D33J013410.7+304635.8 &  2.3354 &   19.43 &   19.57 & 0.48 & 0.32 & 61.28 & 0.00 &     \\ 
 D33J013347.6+304245.3 &  2.3616 &   21.40 &   21.33 & 0.25 & 0.15 & 77.85 & 0.08 &     \\ 
 D33J013416.1+304534.4 &  2.3873 &   21.78 &   21.65 & 0.33 & 0.31 & 79.90 & 0.00 &     \\ 
 D33J013352.6+304713.8 &  2.4157 &   20.69 &   20.73 & 0.57 & 0.38 & 52.68 & 0.00 &     \\ 
 D33J013412.4+304211.1 &  2.5433 &   21.50 &   21.11 & 0.36 & 0.26 & 70.85 & 0.02 &     \\ 
 D33J013405.9+303952.5 &  2.5678 &   20.55 &   20.60 & 0.56 & 0.37 & 65.58 & 0.00 &     \\ 
 D33J013402.3+304409.6 &  2.6614 &   19.49 &   19.66 & 0.47 & 0.36 & 64.42 & 0.00 & (1) \\ 
 D33J013354.3+304029.6 &  2.7104 &   19.12 &   19.21 & 0.62 & 0.38 & 60.52 & 0.01 & (1) \\ 
 D33J013342.1+304255.3 &  2.7169 &   21.08 &   20.96 & 0.31 & 0.31 & 68.40 & 0.04 &     \\ 
 D33J013412.4+304337.9 &  2.7724 & \nodata &   20.25 & 0.68 & 0.32 & 60.43 & 0.02 & (1) \\ 
 D33J013355.6+304053.6 &  2.9579 &   20.29 &   20.25 & 0.41 & 0.24 & 66.10 & 0.02 &     \\ 
 D33J013352.0+303905.0 &  2.9684 &   19.08 &   19.23 & 0.70 & 0.30 & 68.64 & 0.02 & (1) \\ 
 D33J013401.0+304144.9 &  3.3907 &   19.78 &   20.05 & 0.46 & 0.34 & 60.38 & 0.00 &     \\ 
 D33J013352.0+304312.4 &  3.4623 &   20.23 &   20.21 & 0.50 & 0.50 & 56.63 & 0.00 & (1) \\ 
 D33J013409.3+304142.1 &  3.5901 &   20.59 &   20.59 & 0.64 & 0.36 & 47.33 & 0.00 &     \\ 
 D33J013400.8+304623.5 &  3.7015 &   21.17 &   21.26 & 0.63 & 0.37 & 64.12 & 0.01 &     \\ 
 D33J013418.8+304422.7 &  3.8052 &   21.12 &   21.24 & 0.44 & 0.32 & 77.96 & 0.01 &     \\ 
 D33J013356.5+304018.4 &  3.8364 &   20.11 &   20.28 & 0.57 & 0.34 & 68.13 & 0.01 & (1) \\ 
\enddata
\end{planotable}
\end{small}

\begin{small}
\tablenum{2}
\begin{planotable}{lrrrrrcrl}
\tablewidth{40pc}
\tablecaption{\sc Continued}
\tablehead{
\colhead{Name} & \colhead{$P$} & \colhead{} & \colhead{} & \colhead{} &
\colhead{} & \colhead{$i$} & \colhead{} & \colhead{} \\
\colhead{} & \colhead{$(days)$} & \colhead{$V_{max}$} & \colhead{$B_{max}$} &
\colhead{$R_1$} & \colhead{$R_2$} & \colhead{(deg)} & \colhead{$e$} &
\colhead{Comments}}
\startdata
 D33J013349.3+304037.1 &  3.8427 &   19.54 & \nodata & 0.49 & 0.32 & 65.81 & 0.00 & (1) \\ 
 D33J013405.9+304321.4 &  4.1309 &   19.60 &   19.79 & 0.47 & 0.24 & 57.59 & 0.01 &     \\ 
 D33J013407.3+304635.5 &  4.4172 &   19.61 &   19.77 & 0.59 & 0.40 & 61.07 & 0.01 & (1) \\ 
 D33J013347.9+304301.6 &  4.4363 &   19.63 &   19.75 & 0.46 & 0.36 & 71.74 & 0.01 & (1) \\ 
 D33J013350.0+304342.8 &  4.4808 &   20.60 &   20.71 & 0.34 & 0.21 & 89.87 & 0.01 &     \\ 
 D33J013409.1+304112.7 &  4.7310 &   20.72 &   20.92 & 0.31 & 0.25 & 74.58 & 0.10 &     \\ 
 D33J013353.1+303919.1 &  4.7424 &   19.77 &   19.45 & 0.63 & 0.37 & 69.68 & 0.00 &     \\ 
 D33J013426.7+304334.8 &  6.2106 & \nodata &   19.81 & 0.59 & 0.41 & 53.11 & 0.03 & (1) \\ 
 D33J013341.1+304501.7 &  6.8336 &   19.60 &   19.57 & 0.57 & 0.43 & 70.95 & 0.03 & (1) \\ 
 D33J013346.7+304212.3 &  6.9221 &   19.55 &   19.58 & 0.48 & 0.41 & 65.75 & 0.06 & (1) \\ 
 D33J013348.0+304449.6 &  7.2528 &   18.97 &   18.95 & 0.61 & 0.38 & 42.51 & 0.00 & (1) \\ 
 D33J013404.9+304318.4 &  7.2917 &   20.22 &   20.18 & 0.69 & 0.30 & 89.87 & 0.00 & (1) \\ 
 D33J013342.1+304729.7 & 22.9536 &   18.53 &   18.36 & 0.53 & 0.46 & 57.77 & 0.13 & (1) \\ 
\enddata
\tablecomments{(1) Variables identified by Macri et al. (2001a)}
\label{tab:ecl}
\end{planotable}
\end{small}

\begin{small}
\tablenum{3}
\begin{planotable}{ll}
\tablewidth{15pc}
\tablecaption{\sc DIRECT Flux Eclipsing Binaries in M33A}
\tablehead{
\colhead{Name} & \colhead{$P$} \\
\colhead{} & \colhead{$(days)$} }
\startdata
 D33J013358.6+304837.0 &  1.1098 \\      
 D33J013351.9+304022.7 &  1.3107 \\      
 D33J013349.9+304510.7 &  2.1464 \\      
 D33J013345.9+304434.7 &  2.3016 \\      
 D33J013352.7+303929.4 &  2.4432 \\      
 D33J013401.9+303852.9 &  2.6338 \\      
 D33J013406.3+304145.9 &  3.0542 \\      
 D33J013355.4+304522.5 &  3.3443 \\      
 D33J013350.1+304004.6 &  3.3573 \\      
 D33J013408.7+303923.6 &  4.3455 \\      
\enddata   
\label{tab:ecl_flux}
\end{planotable}
\end{small}

\begin{small}
\tablenum{4}
\begin{planotable}{lrrrrrl}
\tablewidth{35pc}
\tablecaption{\sc DIRECT Cepheids in M33A}
\tablehead{
\colhead{Name} & \colhead{$P$} & \colhead{} & \colhead{} & \colhead{} &
\colhead{} & \colhead{} \\
\colhead{} & \colhead{$(days)$} & \colhead{$\langle V\rangle$} &
\colhead{$\langle B\rangle$} & \colhead{$A_V$} & \colhead{$B_V$} &
\colhead{Comments}}
\startdata
 D33J013402.1+304534.0 &  2.823 & \nodata &   22.02 & \nodata &    0.22 &     \\ 
 D33J013342.5+304335.4 &  3.077 &   20.85 &   21.36 &    0.07 &    0.19 &     \\ 
 D33J013416.8+304500.5 &  3.097 &   21.89 &   22.69 &    0.33 &    0.63 &     \\ 
 D33J013407.3+303909.7 &  3.234 &   21.07 &   21.76 &    0.06 &    0.09 &     \\ 
 D33J013420.6+304231.3 &  3.241 &   21.51 &   22.01 &    0.12 &    0.19 &     \\ 
 D33J013358.0+304146.4 &  3.292 &   21.16 &   21.86 &    0.07 &    0.22 &     \\ 
 D33J013350.6+304712.9 &  3.293 &   22.36 &   22.60 &    0.25 &    0.15 &     \\ 
 D33J013405.8+304040.2 &  3.304 &   21.08 &   21.69 &    0.13 &    0.20 &     \\ 
 D33J013427.1+304425.0 &  3.364 &   22.20 &   22.50 &    0.21 &    0.17 &     \\ 
 D33J013409.9+304810.6 &  3.371 &   22.26 & \nodata &    0.23 & \nodata &     \\ 
 D33J013406.5+304420.5 &  3.381 &   21.18 &   21.75 &    0.11 &    0.19 &     \\ 
 D33J013354.2+304624.3 &  3.421 &   21.34 & \nodata &    0.17 & \nodata &     \\ 
 D33J013420.4+304719.9 &  3.443 &   21.79 &   22.00 &    0.20 &    0.22 &     \\ 
 D33J013412.0+304020.2 &  3.478 &   21.22 &   21.72 &    0.10 &    0.14 &     \\ 
 D33J013345.6+304637.9 &  3.563 &   21.02 &   21.60 &    0.07 &    0.11 &     \\ 
 D33J013409.1+304719.0 &  3.569 &   22.72 & \nodata &    0.57 & \nodata &     \\ 
 D33J013356.7+304110.5 &  3.641 &   21.03 &   21.79 &    0.11 &    0.21 &     \\ 
 D33J013427.4+304116.6 &  3.645 & \nodata &   22.50 & \nodata &    0.32 &     \\ 
 D33J013422.9+304220.6 &  3.646 &   21.15 &   21.68 &    0.09 &    0.13 &     \\ 
 D33J013417.3+304505.3 &  3.654 &   22.27 &   22.86 &    0.55 &    0.72 & (1) \\ 
 D33J013343.8+304711.2 &  3.659 &   21.98 &   22.52 &    0.09 &    0.14 &     \\ 
 D33J013359.5+304017.0 &  3.662 &   20.76 &   21.21 &    0.10 &    0.14 &     \\ 
 D33J013347.0+304217.9 &  3.680 &   22.11 &   22.80 &    0.16 &    0.26 &     \\ 
 D33J013420.9+304319.4 &  3.719 &   22.05 &   22.72 &    0.27 &    0.40 &     \\ 
 D33J013354.5+304454.9 &  3.735 &   21.89 &   22.34 &    0.16 &    0.17 &     \\ 
 D33J013339.5+304726.4 &  3.775 &   22.13 &   22.93 &    0.30 &    0.15 &     \\ 
 D33J013407.7+304648.4 &  3.788 &   21.73 &   22.49 &    0.26 &    0.42 &     \\ 
 D33J013410.5+304741.2 &  3.818 &   21.10 &   21.81 &    0.08 &    0.16 &     \\ 
 D33J013348.1+304712.8 &  3.826 &   21.45 &   21.95 &    0.14 &    0.19 &     \\ 
 D33J013351.4+304505.2 &  3.851 &   21.16 &   21.70 &    0.08 &    0.06 &     \\ 
 D33J013414.1+304559.1 &  3.852 &   22.62 &   23.30 &    0.32 &    0.62 &     \\ 
 D33J013353.9+304507.5 &  3.888 &   22.04 &   23.09 &    0.24 &    0.72 &     \\ 
 D33J013417.3+304751.5 &  3.890 &   21.60 &   22.22 &    0.29 &    0.34 &     \\ 
 D33J013405.3+304725.8 &  3.892 &   21.98 & \nodata &    0.20 & \nodata &     \\ 
 D33J013404.4+304319.4 &  4.006 &   21.16 &   21.68 &    0.13 &    0.16 &     \\ 
 D33J013410.3+304103.3 &  4.045 &   22.24 &   22.94 &    0.34 &    0.44 &     \\ 
 D33J013406.0+303945.0 &  4.048 &   20.81 &   21.28 &    0.06 &    0.10 &     \\ 
 D33J013355.9+304507.5 &  4.053 &   21.45 &   21.95 &    0.14 &    0.19 &     \\ 
 D33J013407.6+304626.5 &  4.057 &   21.33 &   21.50 &    0.12 &    0.10 &     \\ 
 D33J013411.4+304655.3 &  4.058 &   21.40 &   21.77 &    0.13 &    0.17 &     \\ 
\enddata
\end{planotable}
\end{small}

\begin{small}
\tablenum{4}
\begin{planotable}{lrrrrrl}
\tablewidth{35pc}
\tablecaption{\sc continued}
\tablehead{
\colhead{Name} & \colhead{$P$} & \colhead{} & \colhead{} & \colhead{} &
\colhead{} & \colhead{} \\
\colhead{} & \colhead{$(days)$} & \colhead{$\langle V\rangle$} &
\colhead{$\langle B\rangle$} & \colhead{$A_V$} & \colhead{$B_V$} &
\colhead{Comments}}
\startdata
 D33J013351.8+304421.3 &  4.093 &   20.99 &   21.65 &    0.18 &    0.18 &     \\ 
 D33J013413.1+304210.5 &  4.110 &   21.69 &   22.16 &    0.25 &    0.49 & (1) \\ 
 D33J013350.7+304512.1 &  4.158 &   21.46 &   22.09 &    0.07 &    0.15 &     \\ 
 D33J013402.3+303855.0 &  4.193 &   21.21 &   21.84 &    0.11 &    0.23 &     \\ 
 D33J013415.6+304004.4 &  4.197 &   21.73 &   22.52 &    0.16 &    0.30 &     \\ 
 D33J013345.1+304229.4 &  4.219 & \nodata &   22.25 & \nodata &    0.22 &     \\ 
 D33J013353.7+304713.5 &  4.240 &   21.94 &   22.72 &    0.28 &    0.44 &     \\ 
 D33J013412.5+304720.7 &  4.261 &   21.59 &   22.24 &    0.20 &    0.42 &     \\ 
 D33J013411.4+303916.2 &  4.301 &   22.06 &   23.75 &    0.17 &    0.68 &     \\ 
 D33J013353.5+304744.1 &  4.312 &   21.20 &   21.64 &    0.19 &    0.22 &     \\ 
 D33J013343.2+304224.9 &  4.337 &   20.91 &   21.33 &    0.12 &    0.24 &     \\ 
 D33J013346.0+304637.0 &  4.346 &   21.96 &   22.54 &    0.25 &    0.23 &     \\ 
 D33J013348.8+304422.1 &  4.359 &   21.17 &   21.84 &    0.09 &    0.15 &     \\ 
 D33J013351.9+304658.7 &  4.391 &   21.41 &   22.12 &    0.35 &    0.54 & (1) \\ 
 D33J013427.7+303954.9 &  4.434 &   22.11 &   22.81 &    0.24 &    0.17 &     \\ 
 D33J013349.0+304040.4 &  4.441 &   21.09 &   21.80 &    0.08 &    0.12 &     \\ 
 D33J013402.6+303944.5 &  4.459 & \nodata &   21.77 & \nodata &    0.30 & (1) \\ 
 D33J013412.2+304353.1 &  4.470 &   21.74 &   22.36 &    0.28 &    0.36 & (1) \\ 
 D33J013356.2+303916.8 &  4.541 &   21.20 &   21.66 &    0.12 &    0.20 &     \\ 
 D33J013356.0+304720.8 &  4.551 &   22.08 & \nodata &    0.39 & \nodata &     \\ 
 D33J013350.9+303918.3 &  4.568 &   21.20 &   21.33 &    0.35 &    0.18 &     \\ 
 D33J013401.2+304813.7 &  4.624 &   22.17 &   22.99 &    0.32 &    0.53 &     \\ 
 D33J013415.7+304524.3 &  4.650 &   21.60 &   22.30 &    0.21 &    0.33 &     \\ 
 D33J013351.7+304548.1 &  4.661 & \nodata &   22.33 & \nodata &    0.23 &     \\ 
 D33J013409.4+304036.8 &  4.694 &   21.41 &   22.03 &    0.15 &    0.23 &     \\ 
 D33J013355.0+304201.6 &  4.702 &   21.35 &   21.99 &    0.15 &    0.16 &     \\ 
 D33J013341.0+304049.6 &  4.707 &   22.09 &   23.16 &    0.22 &    0.09 &     \\ 
 D33J013402.0+304143.5 &  4.713 &   21.80 &   22.48 &    0.32 &    0.56 &     \\ 
 D33J013407.0+304258.5 &  4.725 &   21.44 &   22.19 &    0.21 &    0.35 &     \\ 
 D33J013421.6+304618.2 &  4.767 &   21.70 &   22.45 &    0.32 &    0.52 & (1) \\ 
 D33J013409.7+304350.4 &  4.768 &   21.16 &   21.68 &    0.30 &    0.45 & (1) \\ 
 D33J013402.2+304620.8 &  4.774 &   21.90 &   22.54 &    0.18 &    0.31 &     \\ 
 D33J013351.1+304516.0 &  4.790 &   21.18 &   21.70 &    0.24 &    0.34 &     \\ 
 D33J013359.4+304214.2 &  4.790 &   21.55 &   22.36 &    0.28 &    0.40 & (1) \\ 
 D33J013423.2+304459.3 &  4.803 &   21.94 &   22.77 &    0.29 &    0.62 &     \\ 
 D33J013347.8+304627.4 &  4.858 &   20.85 &   21.43 &    0.11 &    0.15 &     \\ 
 D33J013426.8+304522.0 &  4.869 &   21.44 &   22.29 &    0.35 &    0.68 & (1) \\ 
 D33J013403.0+304428.1 &  4.874 &   22.15 &   22.86 &    0.55 &    0.93 &     \\ 
 D33J013412.2+304620.4 &  4.883 &   21.34 &   22.15 &    0.21 &    0.41 &     \\ 
 D33J013419.0+304604.6 &  4.916 &   21.62 &   22.17 &    0.23 &    0.27 &     \\ 
\enddata
\end{planotable}
\end{small}

\begin{small}
\tablenum{4}
\begin{planotable}{lrrrrrl}
\tablewidth{35pc}
\tablecaption{\sc continued}
\tablehead{
\colhead{Name} & \colhead{$P$} & \colhead{} & \colhead{} & \colhead{} &
\colhead{} & \colhead{} \\
\colhead{} & \colhead{$(days)$} & \colhead{$\langle V\rangle$} &
\colhead{$\langle B\rangle$} & \colhead{$A_V$} & \colhead{$B_V$} &
\colhead{Comments}}
\startdata
 D33J013359.9+304359.3 &  4.969 &   21.42 &   22.02 &    0.21 &    0.29 &     \\ 
 D33J013400.9+304637.9 &  4.987 &   21.68 &   22.57 &    0.14 &    0.24 &     \\ 
 D33J013349.2+304421.7 &  4.990 &   20.95 &   21.43 &    0.07 &    0.14 &     \\ 
 D33J013357.0+304826.4 &  5.008 &   21.69 &   22.28 &    0.23 &    0.38 & (1) \\ 
 D33J013359.4+304338.4 &  5.008 &   21.56 &   22.42 &    0.30 &    0.56 &     \\ 
 D33J013421.5+304000.2 &  5.033 &   21.41 &   22.25 &    0.15 &    0.34 & (1) \\ 
 D33J013352.4+304603.3 &  5.065 &   21.87 &   22.79 &    0.24 &    0.32 &     \\ 
 D33J013405.9+304124.9 &  5.081 &   21.20 &   22.03 &    0.05 &    0.20 &     \\ 
 D33J013351.7+304843.1 &  5.087 &   20.89 & \nodata &    0.12 & \nodata &     \\ 
 D33J013418.3+304602.4 &  5.097 &   21.77 &   22.40 &    0.24 &    0.37 &     \\ 
 D33J013400.4+304808.8 &  5.099 &   21.47 &   22.24 &    0.18 &    0.22 & (1) \\ 
 D33J013410.1+304146.0 &  5.123 &   21.79 &   22.48 &    0.10 &    0.21 &     \\ 
 D33J013341.5+304737.7 &  5.164 &   21.78 &   22.63 &    0.28 &    0.33 & (1) \\ 
 D33J013348.6+304340.0 &  5.164 &   20.94 &   21.07 &    0.14 &    0.19 &     \\ 
 D33J013359.2+304149.4 &  5.228 & \nodata &   21.52 & \nodata &    0.22 &     \\ 
 D33J013350.2+304451.1 &  5.271 &   21.21 &   21.83 &    0.16 &    0.25 &     \\ 
 D33J013359.0+304233.8 &  5.291 & \nodata &   21.98 & \nodata &    0.39 &     \\ 
 D33J013421.6+304415.9 &  5.344 &   21.01 &   21.98 &    0.15 &    0.37 & (1) \\ 
 D33J013347.5+304456.2 &  5.348 & \nodata &   20.73 & \nodata &    0.15 &     \\ 
 D33J013347.2+304452.5 &  5.355 &   20.99 &   21.81 &    0.14 &    0.16 &     \\ 
 D33J013405.5+304133.3 &  5.380 &   21.63 &   22.45 &    0.25 &    0.48 & (1) \\ 
 D33J013358.7+304344.8 &  5.428 &   21.12 &   21.47 &    0.22 &    0.25 &     \\ 
 D33J013359.5+303846.8 &  5.450 &   20.98 &   21.66 &    0.10 &    0.24 & (1) \\ 
 D33J013415.4+304452.8 &  5.479 &   20.61 &   21.14 &    0.13 &    0.19 & (1) \\ 
 D33J013355.3+304638.6 &  5.482 &   21.25 &   21.95 &    0.30 &    0.39 & (1) \\ 
 D33J013356.5+304442.4 &  5.484 &   21.49 &   22.24 &    0.35 &    0.55 &     \\ 
 D33J013411.8+304033.0 &  5.494 &   21.46 &   22.57 &    0.20 &    0.45 &     \\ 
 D33J013404.4+303931.3 &  5.524 &   21.83 & \nodata &    0.17 & \nodata &     \\ 
 D33J013356.0+304007.5 &  5.540 &   21.30 & \nodata &    0.28 & \nodata &     \\ 
 D33J013350.0+304346.7 &  5.572 & \nodata &   22.29 & \nodata &    0.35 & (1) \\ 
 D33J013347.0+304412.3 &  5.597 &   21.28 & \nodata &    0.14 & \nodata &     \\ 
 D33J013346.2+304221.2 &  5.617 &   21.55 &   22.20 &    0.17 &    0.32 &     \\ 
 D33J013402.5+304643.6 &  5.623 &   22.24 &   23.23 &    0.33 &    0.62 &     \\ 
 D33J013356.5+304714.7 &  5.626 &   21.40 &   22.02 &    0.16 &    0.36 &     \\ 
 D33J013413.7+304555.0 &  5.641 &   21.33 &   22.02 &    0.07 &    0.17 &     \\ 
 D33J013414.8+304748.4 &  5.644 &   21.35 &   21.76 &    0.19 &    0.23 &     \\ 
 D33J013414.0+304837.7 &  5.669 &   21.43 &   22.03 &    0.38 &    0.52 & (1) \\ 
 D33J013401.3+304325.6 &  5.670 &   21.14 &   21.77 &    0.40 &    0.58 & (1) \\ 
 D33J013356.5+304428.2 &  5.671 &   21.36 & \nodata &    0.34 & \nodata &     \\ 
 D33J013354.8+304447.0 &  5.700 &   21.07 &   21.88 &    0.12 &    0.21 &     \\ 
\enddata
\end{planotable}
\end{small}

\begin{small}
\tablenum{4}
\begin{planotable}{lrrrrrl}
\tablewidth{35pc}
\tablecaption{\sc continued}
\tablehead{
\colhead{Name} & \colhead{$P$} & \colhead{} & \colhead{} & \colhead{} &
\colhead{} & \colhead{} \\
\colhead{} & \colhead{$(days)$} & \colhead{$\langle V\rangle$} &
\colhead{$\langle B\rangle$} & \colhead{$A_V$} & \colhead{$B_V$} &
\colhead{Comments}}
\startdata
 D33J013420.6+304244.2 &  5.700 &   21.38 &   22.15 &    0.25 &    0.56 & (1) \\ 
 D33J013342.8+304403.1 &  5.721 & \nodata &   20.93 & \nodata &    0.07 &     \\ 
 D33J013403.4+304047.9 &  5.722 &   21.70 &   22.20 &    0.08 &    0.38 &     \\ 
 D33J013406.2+304224.6 &  5.722 &   21.38 &   22.11 &    0.21 &    0.31 &     \\ 
 D33J013350.6+304734.9 &  5.745 &   21.67 &   22.41 &    0.23 &    0.42 & (1) \\ 
 D33J013426.8+304357.7 &  5.752 &   20.09 &   20.29 &    0.13 &    0.18 & (1) \\ 
 D33J013353.1+304542.3 &  5.779 &   22.15 &   23.22 &    0.24 &    0.60 &     \\ 
 D33J013354.8+304518.8 &  5.782 &   20.85 &   21.21 &    0.12 &    0.14 &     \\ 
 D33J013355.7+303903.6 &  5.803 &   21.17 & \nodata &    0.10 & \nodata &     \\ 
 D33J013408.3+304748.0 &  5.811 &   21.70 &   22.59 &    0.33 &    0.54 &     \\ 
 D33J013357.4+304556.6 &  5.817 &   22.05 & \nodata &    0.22 & \nodata &     \\ 
 D33J013420.2+304503.2 &  5.844 &   20.59 &   21.38 &    0.16 &    0.28 &     \\ 
 D33J013349.8+304427.9 &  5.875 &   21.65 &   22.04 &    0.25 &    0.45 & (1) \\ 
 D33J013424.9+304431.2 &  5.921 &   21.30 &   22.21 &    0.21 &    0.44 & (1) \\ 
 D33J013411.9+304704.6 &  5.930 &   22.36 &   23.31 &    0.45 &    0.91 &     \\ 
 D33J013355.6+304449.1 &  5.993 &   22.30 &   23.24 &    0.51 &    0.90 &     \\ 
 D33J013346.9+304334.2 &  5.997 &   21.01 &   21.52 &    0.21 &    0.17 & (1) \\ 
 D33J013349.6+304744.7 &  6.000 &   21.67 &   22.51 &    0.26 &    0.39 & (1) \\ 
 D33J013408.5+304430.6 &  6.000 &   21.50 &   22.10 &    0.23 &    0.28 & (1) \\ 
 D33J013403.2+304332.0 &  6.027 &   21.26 & \nodata &    0.25 & \nodata & (1) \\ 
 D33J013359.0+304531.0 &  6.070 &   21.37 &   22.08 &    0.42 &    0.78 & (1) \\ 
 D33J013356.7+304838.6 &  6.116 &   21.14 &   21.62 &    0.37 &    0.47 & (1) \\ 
 D33J013356.7+304409.5 &  6.140 &   21.44 &   22.30 &    0.31 &    0.58 &     \\ 
 D33J013342.1+304047.3 &  6.178 & \nodata &   22.46 & \nodata &    0.48 &     \\ 
 D33J013406.3+303932.6 &  6.178 &   20.39 &   20.68 &    0.08 &    0.06 &     \\ 
 D33J013403.7+304528.6 &  6.190 &   21.25 &   21.88 &    0.35 &    0.61 & (1) \\ 
 D33J013349.7+304401.6 &  6.280 &   21.47 &   22.22 &    0.29 &    0.40 &     \\ 
 D33J013400.1+303904.2 &  6.292 &   21.16 &   21.90 &    0.24 &    0.35 &     \\ 
 D33J013348.0+304116.8 &  6.298 &   21.09 &   21.64 &    0.17 &    0.09 &     \\ 
 D33J013346.6+304741.7 &  6.319 &   21.19 &   21.88 &    0.15 &    0.14 &     \\ 
 D33J013356.8+304212.2 &  6.329 &   21.17 &   22.11 &    0.23 &    0.39 &     \\ 
 D33J013349.4+304701.9 &  6.335 &   21.16 &   22.05 &    0.35 &    0.47 & (1) \\ 
 D33J013401.9+304148.9 &  6.361 &   21.47 & \nodata &    0.26 & \nodata &     \\ 
 D33J013403.1+303925.9 &  6.376 &   21.82 &   22.78 &    0.35 &    0.74 &     \\ 
 D33J013424.2+304527.5 &  6.432 &   21.44 &   21.97 &    0.21 &    0.27 &     \\ 
 D33J013410.9+303845.1 &  6.478 &   21.32 &   22.48 &    0.32 &    0.49 &     \\ 
 D33J013349.8+304444.7 &  6.519 &   22.03 &   22.93 &    0.32 &    0.47 &     \\ 
 D33J013415.6+304411.9 &  6.529 &   21.50 &   22.32 &    0.34 &    0.61 &     \\ 
 D33J013425.0+304129.7 &  6.550 &   21.29 &   22.06 &    0.41 &    0.76 & (1) \\ 
 D33J013418.5+304118.1 &  6.553 &   20.75 &   21.46 &    0.10 &    0.12 &     \\ 
\enddata
\end{planotable}
\end{small}

\begin{small}
\tablenum{4}
\begin{planotable}{lrrrrrl}
\tablewidth{35pc}
\tablecaption{\sc continued}
\tablehead{
\colhead{Name} & \colhead{$P$} & \colhead{} & \colhead{} & \colhead{} &
\colhead{} & \colhead{} \\
\colhead{} & \colhead{$(days)$} & \colhead{$\langle V\rangle$} &
\colhead{$\langle B\rangle$} & \colhead{$A_V$} & \colhead{$B_V$} &
\colhead{Comments}}
\startdata
 D33J013349.4+304435.5 &  6.842 &   20.82 &   21.57 &    0.04 &    0.14 &     \\ 
 D33J013354.1+304028.1 &  6.862 &   21.61 &   22.73 &    0.21 &    0.29 &     \\ 
 D33J013352.1+304702.9 &  6.870 &   20.63 &   21.10 &    0.14 &    0.26 & (1) \\ 
 D33J013355.8+304416.5 &  6.890 &   21.38 &   21.90 &    0.33 &    0.51 & (1) \\ 
 D33J013353.5+303915.3 &  6.926 &   20.24 & \nodata &    0.29 & \nodata &     \\ 
 D33J013406.4+304003.7 &  6.937 &   20.62 & \nodata &    0.20 & \nodata & (1) \\ 
 D33J013413.5+304704.1 &  6.970 &   21.61 &   22.53 &    0.27 &    0.43 &     \\ 
 D33J013411.6+304742.5 &  6.992 &   20.41 &   21.59 &    0.10 &    0.26 &     \\ 
 D33J013403.0+304727.9 &  7.087 &   21.19 &   21.78 &    0.26 &    0.36 & (1) \\ 
 D33J013404.1+304329.3 &  7.095 &   20.44 &   20.73 &    0.20 &    0.20 & (1) \\ 
 D33J013410.3+303934.8 &  7.121 &   21.39 &   22.30 &    0.20 &    0.28 & (1) \\ 
 D33J013354.1+304504.0 &  7.177 &   21.12 &   21.84 &    0.17 &    0.28 &     \\ 
 D33J013356.5+304644.6 &  7.260 &   21.11 &   21.94 &    0.19 &    0.38 &     \\ 
 D33J013340.7+304543.1 &  7.355 &   21.44 &   22.18 &    0.25 &    0.43 &     \\ 
 D33J013343.2+304502.1 &  7.356 &   20.17 &   20.49 &    0.10 &    0.10 &     \\ 
 D33J013348.6+304416.8 &  7.397 &   22.09 & \nodata &    0.51 & \nodata &     \\ 
 D33J013404.3+304115.4 &  7.455 &   21.42 &   21.94 &    0.20 &    0.45 &     \\ 
 D33J013406.4+303925.6 &  7.510 &   21.21 & \nodata &    0.13 & \nodata &     \\ 
 D33J013347.3+304456.5 &  7.540 &   20.58 &   20.75 &    0.13 &    0.09 &     \\ 
 D33J013405.3+303943.8 &  7.565 &   22.10 &   23.03 &    0.17 &    0.64 &     \\ 
 D33J013402.9+303907.6 &  7.646 &   21.24 &   22.28 &    0.14 &    0.28 &     \\ 
 D33J013349.1+304629.4 &  7.647 &   21.25 &   22.02 &    0.12 &    0.32 &     \\ 
 D33J013350.8+304108.6 &  7.656 &   21.28 &   21.95 &    0.27 &    0.44 &     \\ 
 D33J013422.5+304408.4 &  7.660 &   21.23 &   22.02 &    0.25 &    0.35 & (1) \\ 
 D33J013424.4+304739.0 &  7.690 &   20.90 &   21.57 &    0.35 &    0.59 & (1) \\ 
 D33J013402.3+304242.9 &  7.840 &   20.77 &   21.42 &    0.26 &    0.45 & (1) \\ 
 D33J013344.6+304212.9 &  7.880 &   20.39 &   21.10 &    0.12 &    0.19 & (1) \\ 
 D33J013346.0+304214.6 &  7.894 &   21.30 &   22.40 &    0.28 &    0.27 &     \\ 
 D33J013413.9+304324.3 &  7.950 &   21.42 &   22.40 &    0.25 &    0.44 & (1) \\ 
 D33J013341.8+304312.2 &  8.060 &   21.26 &   22.27 &    0.14 &    0.49 & (1) \\ 
 D33J013355.3+304130.4 &  8.134 &   20.49 &   20.73 &    0.09 &    0.06 &     \\ 
 D33J013417.7+304508.1 &  8.140 &   20.46 &   21.15 &    0.27 &    0.38 & (1) \\ 
 D33J013404.3+304056.0 &  8.220 &   21.57 & \nodata &    0.19 & \nodata &     \\ 
 D33J013412.6+304126.7 &  8.233 &   21.15 &   21.97 &    0.15 &    0.25 & (1) \\ 
 D33J013400.6+304027.8 &  8.257 & \nodata &   20.32 & \nodata &    0.12 &     \\ 
 D33J013356.3+304008.6 &  8.328 &   20.58 &   21.28 &    0.14 &    0.11 &     \\ 
 D33J013342.2+304344.5 &  8.447 &   20.93 &   21.82 &    0.09 &    0.23 &     \\ 
 D33J013411.7+304259.5 &  8.523 & \nodata &   21.22 & \nodata &    0.14 &     \\ 
 D33J013413.3+304307.3 &  8.539 &   21.43 &   22.23 &    0.34 &    0.40 & (1) \\ 
 D33J013416.0+304644.0 &  8.560 &   21.14 &   21.93 &    0.23 &    0.37 & (1) \\ 
\enddata
\end{planotable}
\end{small}

\begin{small}
\tablenum{4}
\begin{planotable}{lrrrrrl}
\tablewidth{35pc}
\tablecaption{\sc continued}
\tablehead{
\colhead{Name} & \colhead{$P$} & \colhead{} & \colhead{} & \colhead{} &
\colhead{} & \colhead{} \\
\colhead{} & \colhead{$(days)$} & \colhead{$\langle V\rangle$} &
\colhead{$\langle B\rangle$} & \colhead{$A_V$} & \colhead{$B_V$} &
\colhead{Comments}}
\startdata
 D33J013356.2+303909.1 &  8.640 &   20.68 &   21.27 &    0.15 &    0.21 & (1) \\ 
 D33J013409.3+304238.6 &  8.714 &   21.10 &   22.05 &    0.21 &    0.49 & (1) \\ 
 D33J013413.5+304334.7 &  9.098 &   20.93 &   21.57 &    0.30 &    0.46 & (1) \\ 
 D33J013342.9+304305.5 &  9.251 &   20.85 &   21.43 &    0.17 &    0.07 &     \\ 
 D33J013426.9+304003.6 &  9.260 &   20.89 &   21.65 &    0.24 &    0.23 & (1) \\ 
 D33J013341.1+304340.7 &  9.316 &   20.96 &   21.70 &    0.13 &    0.15 &     \\ 
 D33J013345.1+304021.8 &  9.388 &   20.80 &   21.70 &    0.17 &    0.22 &     \\ 
 D33J013349.6+304501.0 &  9.582 &   21.17 &   21.96 &    0.19 &    0.21 &     \\ 
 D33J013353.2+304835.2 &  9.600 &   20.42 & \nodata &    0.14 & \nodata & (1) \\ 
 D33J013350.8+304715.5 &  9.710 &   20.77 &   21.41 &    0.23 &    0.42 & (1) \\ 
 D33J013421.1+304415.5 &  9.980 &   20.72 &   21.36 &    0.29 &    0.46 & (1) \\ 
 D33J013408.8+303946.5 & 10.110 &   20.45 &   21.30 &    0.11 &    0.28 & (1) \\ 
 D33J013343.2+304002.7 & 10.376 &   20.87 &   21.63 &    0.18 &    0.21 &     \\ 
 D33J013346.7+304445.6 & 10.383 &   20.29 &   20.67 &    0.12 &    0.03 &     \\ 
 D33J013356.1+303903.0 & 10.430 &   20.71 &   21.64 &    0.17 &    0.17 & (1) \\ 
 D33J013410.2+304450.6 & 10.450 &   20.81 &   21.63 &    0.30 &    0.52 & (1) \\ 
 D33J013417.0+304640.2 & 10.560 &   20.58 &   21.47 &    0.24 &    0.40 & (1) \\ 
 D33J013413.9+304418.7 & 10.610 &   21.36 &   22.29 &    0.44 &    0.70 & (1) \\ 
 D33J013420.0+304302.3 & 10.727 &   21.24 &   22.08 &    0.13 &    0.18 &     \\ 
 D33J013349.5+304101.5 & 10.919 &   20.69 &   21.31 &    0.17 &    0.20 &     \\ 
 D33J013358.6+304400.1 & 11.150 &   20.52 &   21.34 &    0.14 &    0.29 & (1) \\ 
 D33J013340.3+304741.7 & 11.510 &   21.06 &   21.88 &    0.09 &    0.29 & (1) \\ 
 D33J013346.5+304111.8 & 11.511 &   20.51 &   21.25 &    0.07 &    0.05 &     \\ 
 D33J013357.4+304113.9 & 11.620 &   20.37 &   21.20 &    0.22 &    0.41 & (1) \\ 
 D33J013401.6+303858.2 & 11.707 &   20.46 &   21.27 &    0.35 &    0.30 &     \\ 
 D33J013414.7+304609.7 & 11.770 &   21.55 &   22.79 &    0.27 &    0.60 & (1) \\ 
 D33J013356.2+304343.0 & 11.970 &   20.84 &   21.68 &    0.28 &    0.56 & (1) \\ 
 D33J013420.3+304351.9 & 11.970 &   20.69 &   21.50 &    0.40 &    0.61 & (1) \\ 
 D33J013341.4+304756.1 & 12.120 &   20.25 &   21.01 &    0.18 &    0.19 & (1) \\ 
 D33J013346.6+304821.8 & 12.360 &   20.36 &   21.44 &    0.18 &    0.31 & (1) \\ 
 D33J013351.1+304400.4 & 12.360 &   20.89 &   21.81 &    0.32 &    0.67 & (1) \\ 
 D33J013411.3+304755.1 & 12.410 &   20.35 &   21.09 &    0.30 &    0.47 & (1) \\ 
 D33J013355.0+304643.2 & 12.870 &   20.31 &   21.07 &    0.15 &    0.24 & (1) \\ 
 D33J013346.0+304231.9 & 12.930 &   20.14 &   20.80 &    0.21 &    0.21 & (1) \\ 
 D33J013402.8+304145.7 & 13.040 &   20.09 &   20.83 &    0.30 &    0.55 & (1) \\ 
 D33J013345.9+304421.4 & 13.120 &   20.49 &   21.30 &    0.31 &    0.51 & (1) \\ 
 D33J013359.9+303910.3 & 13.230 &   20.66 &   21.55 &    0.19 &    0.37 & (1) \\ 
 D33J013357.3+303840.1 & 13.337 & \nodata &   20.98 & \nodata &    0.14 &     \\ 
 D33J013415.1+304435.2 & 13.340 &   20.69 &   21.73 &    0.24 &    0.41 & (1) \\ 
 D33J013356.5+304632.1 & 13.348 &   21.88 & \nodata &    0.54 & \nodata & (1) Type II\\ 
\enddata
\end{planotable}
\end{small}

\begin{small}
\tablenum{4}
\begin{planotable}{lrrrrrl}
\tablewidth{35pc}
\tablecaption{\sc continued}
\tablehead{
\colhead{Name} & \colhead{$P$} & \colhead{} & \colhead{} & \colhead{} &
\colhead{} & \colhead{} \\
\colhead{} & \colhead{$(days)$} & \colhead{$\langle V\rangle$} &
\colhead{$\langle B\rangle$} & \colhead{$A_V$} & \colhead{$B_V$} &
\colhead{Comments}}
\startdata
 D33J013408.8+304543.5 & 13.530 &   20.85 &   21.63 &    0.37 &    0.58 & (1) \\ 
 D33J013400.9+304028.5 & 13.672 &   19.98 &   20.91 &    0.28 &    0.20 &     \\ 
\enddata 
\tablecomments{(1) Variables identified by Macri et al. (2001a)} 
\label{tab:ceph}
\end{planotable} 
\end{small}

\begin{small}
\tablenum{5}
\begin{planotable}{lrl}
\tablewidth{20pc}
\tablecaption{\sc DIRECT Flux Cepheids in M33A}
\tablehead{
\colhead{Name} & \colhead{$P$} & \colhead{} \\
\colhead{} & \colhead{$(days)$} & \colhead{Comments}}
\startdata
 D33J013341.5+304149.7 &  3.400 &     \\ 
 D33J013352.3+303901.0 &  3.663 &     \\ 
 D33J013346.0+303908.3 &  4.087 &     \\ 
 D33J013353.1+304619.1 &  4.426 &     \\ 
 D33J013344.5+304644.8 &  4.536 &     \\ 
 D33J013405.6+304143.5 &  4.716 &     \\ 
 D33J013357.7+304834.8 &  4.856 &     \\ 
 D33J013350.9+303914.6 &  5.362 &     \\ 
 D33J013351.1+303931.4 &  5.457 &     \\ 
 D33J013344.1+304558.1 &  5.568 & (1) \\ 
 D33J013417.1+303932.9 &  5.650 & (1) \\ 
 D33J013413.6+304210.5 &  5.867 & (1) \\ 
 D33J013343.9+304513.7 &  5.910 & (1) \\ 
 D33J013345.6+303921.5 &  6.008 &     \\ 
 D33J013342.2+304128.3 &  6.559 &     \\ 
 D33J013350.5+304502.0 &  6.763 &     \\ 
 D33J013406.7+304041.2 &  6.766 &     \\ 
 D33J013344.6+303919.9 &  7.452 &     \\ 
 D33J013345.2+304428.2 &  7.599 &     \\ 
 D33J013355.1+304758.5 &  8.242 &     \\ 
 D33J013354.4+304527.2 &  8.747 &     \\ 
 D33J013347.4+303848.5 &  9.066 &     \\ 
 D33J013352.8+303836.6 & 11.168 &     \\ 
 D33J013412.5+303839.8 & 11.240 & (1) \\ 
 D33J013356.0+304231.3 & 11.340 & (1) \\ 
 D33J013420.2+304457.9 & 12.934 &     \\ 
 D33J013408.1+303931.9 & 13.320 & (1) \\ 
\enddata
\tablecomments{(1) Variables identified by Macri et al. (2001a)}
\label{tab:ceph_flux}
\end{planotable}
\end{small}

\begin{small}
\tablenum{6}
\begin{planotable}{lrrrrrl}
\tablewidth{35pc}
\tablecaption{\sc DIRECT Periodic Variables in M33A}
\tablehead{
\colhead{Name} & \colhead{$P$} & \colhead{$V^a$} &
\colhead{$B^a$} & \colhead{} & \colhead{} & \colhead{} \\
\colhead{} & \colhead{$(days)$} & \colhead{} &
\colhead{} & \colhead{$A_V$} & \colhead{$B_V$} &
\colhead{Comments}}
\startdata
 D33J013407.1+303925.6 &  1.58 &   19.51 &   19.30 &    0.01 &    0.02 &     \\ 
 D33J013345.0+304449.3 &  1.65 &   18.76 & \nodata &    0.03 & \nodata & EB  \\ 
 D33J013418.4+304552.2 &  1.97 & \nodata &   21.55 & \nodata &    0.68 & EB  \\ 
 D33J013426.4+304033.3 &  2.24 &   20.70 & \nodata &    0.05 & \nodata & EB  \\ 
 D33J013340.9+303911.0 &  2.45 &   21.13 &   21.38 &    0.28 &    0.32 & EB  \\ 
 D33J013345.9+304748.6 &  2.51 &   21.40 &   21.30 &    0.11 &    0.41 & EB  \\ 
 D33J013427.3+304311.6 &  3.59 & \nodata &   21.87 & \nodata &    0.53 & EB  \\ 
 D33J013349.3+303835.2 &  4.99 &   19.20 &   19.53 &    0.04 &    0.01 &     \\ 
 D33J013344.2+304214.7 &  6.11 &   20.10 &   20.20 &    0.31 &    0.19 & EB  \\ 
 D33J013352.4+303840.2 &  6.38 &   19.90 &   19.75 &    0.06 &    0.03 &     \\ 
 D33J013346.0+304658.5 &  6.46 &   21.21 &   21.72 &    0.27 &    0.57 & EB  \\ 
 D33J013409.2+304641.4 & 11.65 &   18.31 &   18.31 &    0.02 &    0.03 &     \\ 
 D33J013405.5+304726.3 & 13.17 &   18.33 &   18.50 &    0.02 &    0.02 &     \\ 
\enddata
\tablecomments{
$^a$ The $V$ and $B$ columns list the maximum magnitudes $V_{max}$
and $B_{max}$ for the eclipsing variables and flux-weighted average
magnitudes $\langle V\rangle$ and $\langle B\rangle$ for the other
variables.\\
}
\label{tab:per}
\end{planotable}
\end{small}

\begin{small}
\tablenum{7}
\begin{planotable}{lrrrrl}
\tablewidth{35pc}
\tablecaption{\sc DIRECT Miscellaneous Variables in M33A}
\tablehead{
\colhead{Name} & \colhead{$\bar{V}$} &
\colhead{$\bar{B}$} & \colhead{$A_V$} & \colhead{$B_V$} & \colhead{Comments}}
\startdata
 D33J013339.7+304541.6 &   17.27 &   17.25 &    0.46 &    0.20 &      \\ 
 D33J013407.5+304721.4 &   19.24 & \nodata &    0.13 & \nodata &      \\ 
 D33J013412.7+304728.2 &   19.37 & \nodata &    0.23 & \nodata &      \\ 
 D33J013416.8+304518.8 &   19.37 &   21.26 &    0.61 &    0.52 & (1)  \\ 
 D33J013344.8+304249.9 &   19.41 &   20.51 &    0.14 &    0.13 &      \\ 
 D33J013416.6+304750.9 &   19.55 & \nodata &    0.22 & \nodata & (1)  \\ 
 D33J013348.8+304212.0 &   19.62 & \nodata &    0.20 & \nodata & (1)  \\ 
 D33J013345.8+304438.1 &   19.81 & \nodata &    0.17 & \nodata &      \\ 
 D33J013347.5+304630.9 &   19.94 &   21.70 &    0.28 &    0.37 & (1)  \\ 
 D33J013350.5+303836.3 &   19.98 & \nodata &    0.96 & \nodata &      \\ 
 D33J013409.4+304527.3 &   20.00 &   21.62 &    0.24 &    0.34 & (1)  \\ 
 D33J013421.0+304743.0 &   20.18 &   21.67 &    0.30 &    0.53 &      \\ 
 D33J013342.3+304024.7 &   20.27 &   20.26 &    0.11 &    0.07 &      \\ 
 D33J013420.1+304436.9 &   20.40 &   22.07 &    0.39 &    0.47 & (1)  \\ 
 D33J013409.7+304100.2 &   20.49 &   21.91 &    0.52 &    0.64 &      \\ 
 D33J013345.9+304312.4 &   20.50 &   20.97 &    0.37 &    0.34 &      \\ 
 D33J013414.3+304228.7 &   20.57 &   22.90 &    0.34 &    0.47 &      \\ 
 D33J013406.1+304305.4 &   20.60 &   20.61 &    0.29 &    0.22 &      \\ 
 D33J013358.8+303928.1 &   20.68 &   21.18 &    0.43 &    0.37 &      \\ 
 D33J013347.7+304030.9 &   20.69 &   21.34 &    0.40 &    0.39 &      \\ 
 D33J013341.9+304021.1 &   20.89 & \nodata &    0.33 & \nodata &      \\ 
 D33J013410.2+304107.1 &   20.90 &   21.04 &    0.32 &    0.38 &      \\ 
 D33J013413.6+303939.4 &   20.90 &   21.41 &    0.42 &    0.20 &      \\ 
 D33J013400.8+304339.9 &   21.05 &   20.99 &    0.28 &    0.26 &      \\ 
 D33J013359.8+304725.3 &   21.06 &   22.70 &    0.40 &    0.85 &      \\ 
 D33J013349.0+303906.2 &   21.12 & \nodata &    0.87 & \nodata &      \\ 
 D33J013346.7+304210.5 &   21.13 & \nodata &    0.96 & \nodata & (1)  \\ 
 D33J013412.3+304551.0 &   21.16 &   21.96 &    0.41 &    0.38 &      \\ 
 D33J013402.1+304038.3 &   21.19 & \nodata &    0.35 & \nodata &      \\ 
 D33J013349.2+303909.7 &   21.20 &   22.25 &    0.77 &    0.42 &      \\ 
 D33J013419.5+304151.9 &   21.21 & \nodata &    0.38 & \nodata &      \\ 
 D33J013357.0+304044.2 &   21.24 &   22.41 &    0.83 &    0.46 &      \\ 
 D33J013348.5+304015.0 &   21.25 &   21.98 &    0.85 &    0.31 &      \\ 
 D33J013405.3+303954.2 &   21.32 &   22.65 &    0.40 &    0.65 &      \\ 
 D33J013345.1+304715.3 &   21.38 & \nodata &    0.77 & \nodata & (1)  \\ 
 D33J013413.3+303848.2 &   21.38 &   21.57 &    0.60 &    0.37 &      \\ 
 D33J013416.4+304045.9 &   21.41 &   22.50 &    0.78 &    0.34 &      \\ 
 D33J013343.4+304248.4 &   21.43 &   22.61 &    0.66 &    0.43 &      \\ 
 D33J013357.2+304523.0 &   21.47 &   22.25 &    0.43 &    0.66 &      \\ 
 D33J013412.7+304208.4 &   21.52 & \nodata &    0.52 & \nodata & (1)  \\ 
\enddata
\end{planotable}
\end{small}

\begin{small}
\tablenum{7}
\begin{planotable}{lrrrrl}
\tablewidth{35pc}
\tablecaption{\sc continued}
\tablehead{
\colhead{Name} & \colhead{$\bar{V}$} &
\colhead{$\bar{B}$} & \colhead{$A_V$} & \colhead{$B_V$} & \colhead{Comments}}
\startdata
 D33J013414.5+304040.4 &   21.56 & \nodata &    0.59 & \nodata &      \\ 
 D33J013407.7+303845.5 &   21.63 & \nodata &    0.81 & \nodata &      \\ 
 D33J013341.7+304145.7 & \nodata & \nodata & \nodata & \nodata &      \\ 
 D33J013345.5+303912.2 & \nodata & \nodata & \nodata & \nodata &      \\ 
 D33J013346.1+303917.7 & \nodata & \nodata & \nodata & \nodata &      \\ 
 D33J013351.5+304050.9 & \nodata & \nodata & \nodata & \nodata &      \\ 
 D33J013352.4+304501.7 & \nodata & \nodata & \nodata & \nodata &      \\ 
 D33J013357.6+303844.0 & \nodata & \nodata & \nodata & \nodata &      \\ 
 D33J013406.0+304006.0 & \nodata & \nodata & \nodata & \nodata &      \\ 
 D33J013414.2+304740.3 & \nodata &   22.19 & \nodata &    0.52 &      \\ 
\enddata
\tablecomments{(1) Variables identified by Macri et al. (2001a)}
\label{tab:misc}
\end{planotable}
\end{small}

\end{document}